\documentclass[12pt]{article}

\usepackage{float}

\usepackage{titling}
\usepackage{etoolbox}

\renewenvironment{abstract}
 {\normalsize
  \begin{center}
  \bfseries \abstractname\vspace{-.5em}\vspace{0pt}
  \end{center}
  \list{}{%
    \setlength{\leftmargin}{1cm}%
    \setlength{\rightmargin}{\leftmargin}%
  }%
  \item\relax}
 {\endlist}
\newenvironment{titleabstractblock}
  {\begin{samepage}}
  {\end{samepage}}
 
\newcommand{\TITLE}[1]{\begin{center}\normalsize\bfseries #1\end{center}}
\newcommand{\ARTICLEAUTHORS}[1]{\begin{center}\small #1\end{center}\vspace{0.5em}}
\newcommand{\ABSTRACT}[1]{\renewcommand{\abstractname}{}\begin{abstract}#1\end{abstract}}
\newcommand{\KEYWORDS}[1]{\begin{center}\normalsize\textbf{Keywords:} #1\end{center}}

\usepackage[margin=1in]{geometry}  % Set margins

\def\fps@table{H}

\usepackage[margin=1in]{geometry}  % Set margins
\usepackage{setspace}              % For line spacing
\usepackage[table]{xcolor}
\usepackage{graphicx}
\usepackage{caption}
\usepackage{url}
\usepackage[skip=0.5ex]{subcaption}
\usepackage{soul}
\usepackage{placeins}
\usepackage{amsmath}
\usepackage{afterpage}
\usepackage{changes}
\usepackage{tikz}
\usepackage{bm}
\usetikzlibrary{calc}
\usepackage{pgfplots}
\pgfplotsset{compat=1.15}
\usepackage{comment}
\usepackage{multirow}
\usepackage{bbm}
\usepackage{enumitem}
\usepackage{arydshln}
\usepackage{float}
\usepackage{booktabs}

\usepackage{natbib}
\bibpunct[, ]{(}{)}{,}{a}{}{,}

\numberwithin{equation}{section}

\newcolumntype{C}[1]{>{\centering\let\newline\\\arraybackslash\hspace{0pt}}m{#1}}
\newcolumntype{L}[1]{>{\raggedright\let\newline\\\arraybackslash\hspace{0pt}}m{#1}}
\newcolumntype{R}[1]{>{\raggedleft\let\newline\\\arraybackslash\hspace{0pt}}m{#1}}

\newcounter{marginparcounter}

\usepackage{amsthm}

\theoremstyle{definition}

\theoremstyle{remark}

\pretitle{\begin{center}\normalsize\bfseries}
\posttitle{\par\end{center}\vspace{0em}}
\preauthor{\begin{center}\small}
\postauthor{\end{center}\vspace{0em}}
\predate{}
\postdate{}

\begin{document}

% Begin unbreakable title and abstract block
\begin{titleabstractblock}

\TITLE{Deploying Chatbots in Customer Service: \\Adoption Hurdles and Simple Remedies}

\ARTICLEAUTHORS{%
Evgeny Kagan\\
\textit{Carey Business School, Johns Hopkins University}\\[1em]
Maqbool Dada\\
\textit{Carey Business School, Johns Hopkins University}\\[1em]
Brett Hathaway\\
\textit{Marriott School of Business, Brigham Young University}
}

\ABSTRACT{\textbf{Problem definition:} Despite recent advances in Artificial Intelligence, the use of chatbot technology in customer service continues to face adoption hurdles. This paper explores reasons for these adoption hurdles and tests several service design levers to increase chatbot uptake. \textbf{Methodology/results:} We use incentivized online experiments to study chatbot uptake in a variety of scenarios. The results of these experiments are threefold. First, people respond positively to improvements in chatbot performance; however, the chatbot channel is utilized less frequently than expected-time minimization would predict. A key driver of this underutilization is the reluctance to engage with a gatekeeper process, i.e., a process with an imperfect initial service stage and possible transfer to a second, expert service stage -- a behavior we term \textit{gatekeeper aversion}. We show that gatekeeper aversion can be further amplified by a secondary hurdle, algorithm aversion. Second, chatbot uptake can be increased by providing customers with average waiting times in the chatbot channel, as well as by being more transparent about chatbot capabilities and limitations. Third, methodologically, we show that chatbot adoption can depend on experimental implementation. In particular, chatbot adoption decreases further as (i) stakes are increased, (ii) the human/algorithmic nature of the server is manipulated with more realism. \textbf{Managerial Implications:} Our results suggest that firms should continue to prioritize investments in chatbot technology. However, less expensive, process-related interventions can also be effective. These may include being more transparent about the types of queries that are (or are not) suitable for chatbots, emphasizing chatbot reliability and quick resolution times, as well as providing faster live agent access to customers who experienced chatbot failure.}
 

% End unbreakable title and abstract block
\end{titleabstractblock}

\section{Introduction}
Recent technological advances have significantly increased chatbot capabilities, improved their speed, enabled them to handle more complex, often unstructured  customer queries, and reduced training and maintenance costs \citep{johannsen2018}.\footnote{This version of the paper makes references to the Electronic Companion (EC). The EC can be found attached to the SSRN version of the paper, please see \url{https://papers.ssrn.com/sol3/papers.cfm?abstract_id=4283285}.} These improvements have reduced the staffing needs for live operators, lowering payroll and other costs related to providing live customer support. The cost savings can be substantial -- a recent report estimates an average cost reduction of up to \$0.70 per customer interaction, and an annual savings of 8 Billion US Dollars in the banking sector alone \citep{juniper2020}. 

The technological maturity and the cost savings offered by chatbots have shifted the burden of successful chatbot deployment from AI developers to managers implementing this technology in their organizations. However, academic research into the drivers of chatbot technology adoption remains scarce.  While there is a growing literature on human-chatbot interactions  \citep{goot2019,goot2020,sheehan2020,schanke2021,adam2021,benke2022}, it is focused mainly on questions related to chatbot design; for example, on whether anthropomorphism (human-likeness) helps or hurts adoption. These studies help developers build chatbots with more desirable appearance and behavior; however, they provide little or no insight into the process design implications of deploying chatbots, their integration into the broader service delivery strategy, and their effects on the cost and performance of a service system. In this paper we seek to address this gap and study chatbot technology from a service operations perspective. We focus on chatbot adoption as a choice among several service channels offered within a service system, each with its own unique processes and customer experiences.

Operationally, chatbot systems resemble gatekeeper systems \citep{shumsky2003,freeman2017,hathaway2022}, where the chatbot plays the role of a gatekeeper that handles only a subset of the incoming requests, with the remaining requests being diverted to a live, human agent. This is because certain requests may be difficult to communicate or categorize,  or because the chatbot may not be authorized to handle certain requests; for example, ones that involve large financial transactions. Thus, the chatbot serves as the entry point to, but not necessarily the final step of, the service encounter, similar to a nurse in a hospital or a front desk receptionist in a hotel. Different from many healthcare or hospitality settings, 
which \textit{require} the patient or customer to go through the gatekeeper to begin service,  chatbot operators often allow customers to \textit{choose} between a live agent and a chatbot. 
In this study we examine the determinants of this channel choice and test several levers to increase chatbot uptake.

\subsection{Study design}
The starting point of our investigation is a retrospective survey, in which we ask 400 respondents to describe a recent customer service episode, either with a chatbot or with a live agent. Quantitative and qualitative analyses of their testimonies suggest a key trade-off in channel choice: chatbots are faster to access but have a lower request resolution rate. In contrast, live agents typically require some wait to access but are much more reliable in resolving customer requests. This insight helps motivate a simple model of channel choice (\textsection 2) which is then tested in our experiments (\textsection 3-5). 

The first experiment (\textsection\ref{sec:exp:1}) focuses on identifying adoption hurdles. In this experiment we present participants with a series of choices between two alternatives. The first alternative (``Channel A'') represents the live (A)gent and involves some waiting in line to access the server. The server then resolves the request with probability 1. The second alternative  (``Channel B'')  represents the (B)ot and involves no waiting to access the first service stage; however, the server fails with some known probability, requiring additional waiting in line and a second service stage. In both channels, upon successful resolution of the service request the customer receives a fixed monetary reward, which represents service completion. Depending on the parameters in a decision, the expected-time minimizing choice may be either Channel A or Channel B. There are three experimental conditions: a \textit{Context} treatment, in which the type of the server (human or bot) is explicitly revealed, a \textit{No Context} treatment, in which all contextual cues are removed and participants choose between two visually identical (but process-differentiated) channels, and a \textit{No Context, Deterministic} treatment which removes the uncertainty from Channel B. These treatments enable us to separately identify process-related preferences that exist independently of contextual framing from preferences related to the algorithmic nature of the service provider.

In the second experiment (\textsection 4) we focus on potential remedies for chatbot underutilization. Drawing on the literature in behavioral operations and decision theory, we test two alternative designs. In particular, we present participants with choices that are mathematically identical to those in Experiment 1 but change how information is presented. First, drawing on research on operational transparency \citep{buell2011,buell2017,balakrishnan2022}, the \textit{Context + No Transparency} treatment deliberately reduces operational transparency. In this treatment, the chatbot always suggests a solution for each customer request, with the offered solution being either correct or incorrect. This is different from our  \textit{Context} treatment, where the chatbot is transparent and truthfully reports whenever it is able to handle a request or not.  Second, in the \textit{Context + Nudge} treatment we nudge participants to focus on the potential time savings offered by the chatbot by explicitly presenting the total average waiting times for both channels. 

In the third experiment (\textsection\ref{sec:exp:3}) we turn to the methodological challenge of measuring algorithmic aversion in the customer service setting by introducing two treatments that add realism to our experimental setup. In the first treatment (\textit{Context + Live}) we replicate the \textit{Context} treatment but use actual humans (research assistants, blind to the experimental hypotheses) who play the role of live agents and who interact with participants using a live chat tool. In the second treatment (\textit{Context + Hold}) we make salient differences between live agents and chatbots by requiring continuous, physical engagement in Channel A (representing the live agent) while retaining click-based interaction in Channel B (representing the chatbot). 

\subsection{Key Results and Contributions}
The results of our experiments show that Channel B uptake declines as the chatbot service time grows longer, chatbot failure rate increases, or the wait for a live agent following chatbot failure increases. In other words, better operational performance leads to higher uptake. Nonetheless, across all three experiments, Channel B uptake remains considerably below what one would predict from a purely expected-time minimization perspective. Our results are summarized in Table \ref{tab:intro}. %Further, as waiting times go up (all time parameters in both channels are multiplied by a factor of two), Channel B uptake decreases even further, suggesting that chatbot underutilization is more pronounced when stakes are higher.  

In Experiment 1 we first show that Channel B underutilization is tied primarily to process-related hurdles. That is, participants are willing to spend more time in the system (in expectation) in order to avoid interacting with a gatekeeper channel, whether or not the decision is contextualized (as a choice between a live agent and a chatbot) or not. We term this behavior \textit{gatekeeper aversion}. We further decompose gatekeeper aversion into two distinct components: risk aversion (preference for a less uncertain service time duration) and transfer aversion (preference for continuous, rather than fragmented, multi-stage service processes). While risk aversion is well-documented in financial contexts \citep{holt2001,harrison2008}, customer behaviors in the presence of uncertain waiting times and fragmented service processes have received little attention in the behavioral literature \citep{allon2018}. Thus, our first theoretical contribution is to document and characterize this important customer preference. 

Continuing with Experiment 1, we show that algorithmic context may further affect chatbot uptake. The standard result in the literature is that AI assistance is often underutilized, even when AI performs as well as or better than a human alternative  \citep{dietvorst2015,logg2019,castelo2019}. We first follow the standard approach in the literature and conduct a series of pre-tests that hold the processes and performance constant across the two channels and vary only their labels and visual cues. These pre-tests do not detect any algorithm aversion, suggesting that the classic result that algorithmic errors loom larger than human errors does not hold in our setting. Nonetheless, we show that algorithm aversion still matters. Specifically, we show that gatekeeper aversion and algorithm aversion may interact to produce significantly lower chatbot uptake than can be explained by gatekeeper aversion alone, particularly when the stakes (waiting times in both channels) are high. Thus, our second theoretical contribution is to show that algorithm aversion can serve as an \textit{amplifier}, reinforcing the reluctance to use a service channel with a gatekeeper structure. 

In Experiment 2 we show that the aversions identified in Experiment 1 can be mitigated by varying how information is presented to customers. In particular, both transparency about chatbot capabilities and the average waiting time nudge can significantly increase chatbot adoption, although their effectiveness varies with time scale. Specifically, operational transparency matters when durations are short (suggesting that its effect is washed out when stakes are higher), whereas the effect of the nudge is more robust.

These results of Experiments 1 and 2 suggest practical ways to increase chatbot adoption: through fast-tracking chatbot customers by shortening their wait for a live agent after chatbot failure, through communicating operational advantages via explicit, time-based metrics and through providing a candid account of chatbot capabilities, rather than attempting to handle all inquiries without disclosure. In \textsection 4.4 we use a structural estimation of utility parameters to build a counterfactual model of channel-joining behavior and show that our interventions achieve substantial staffing cost savings (of up to 19.7\%) in moderately congested systems. Thus, a practical contribution of our study is to identify inexpensive and easily implementable service design interventions that can increase chatbot adoption and generate substantial cost savings.

In the third experiment we show that increasing the realism of the interactions produces behaviors that are consistent with our original design (with similar aversion magnitudes). However, greater realism introduces a small but significant increase in algorithm aversion compared to the \textit{Context} treatment. Thus, we contribute to the methodological discourse on measuring algorithmic attitudes by showing that experimental designs that rely on contextual framing alone may underestimate algorithm aversion, compared to designs that involve longer interactions or vary the human versus algorithmic nature of the interaction in a more realistic manner.

\begin{center}
\begin{table}[bt]
\caption{Results Summary}
\label{tab:intro}
\renewcommand{\arraystretch}{1.25}
\footnotesize
\begin{tabular}{L{3.2cm} l p{4.5cm} l}
\toprule
 & \multicolumn{2}{l}{\textbf{Effect}} & \textbf{Effect detected?} \\ 
\midrule
 \multicolumn{4}{c}{\textbf{\uline{Experiment 1: Adoption Hurdles (\textsection 3)}}}  \\
H1.1: & Transfer aversion & \multirow{2}{*}{\scalebox{1.6}{\}} \hspace{-0.2em} Gatekeeper  Aversion}& *** \\
H1.2: & Risk aversion & & *** \\
Pre-tests of \textit{Context} manipulation: & \multicolumn{2}{l}{Algorithm aversion (Standalone effect)} & n.s. \\
H1.3: & \multicolumn{2}{l}{Algorithm aversion (Amplifying effect)} & n.s. for short dur.,$~~~$ ** for long dur. \\
 \hdashline
 \addlinespace
 
  \multicolumn{4}{c}{\uline{\textbf{Experiment 2: Remedies (\textsection 4)}}}  \\
H2.1: & \multicolumn{2}{l}{Average Waiting Time Nudge} & *** \\
H2.2: & \multicolumn{2}{l}{Transparency} & ** for short dur., $~~~$ n.s. for long dur. \\
 \hdashline
 \addlinespace
 
  \multicolumn{4}{c}{\uline{\textbf{Experiment 3: Alternative Measurements of Algorithm Aversion (\textsection 5)}}}  \\
H3.1: & \multicolumn{2}{l}{Algorithm aversion (\textit{Context + Live})} & ** \\
H3.2: & \multicolumn{2}{l}{Algorithm aversion (\textit{Context + Hold})} & *** \\
\bottomrule
\end{tabular}
\vspace{0.3em}
\begin{minipage}{\textwidth}
\setlength{\baselineskip}{0.3\baselineskip}
\footnotesize\textit{Note}: Asterisks denote $p-$values after Bonferroni-Holm multiple hypothesis adjustment, following \cite{holm1979} and \cite{list2019}. Effect sizes and standard errors are based on specifications in Tables 4, 5, and 6. *** $p < 0.01$, ** $p < 0.05$, * $p < 0.1$, n.s. = not significant.
\end{minipage}
\end{table}
 \end{center}

\section{Retrospective Survey, Literature and Experiment Design}
To motivate our model and experimental approach, we first report the results of a retrospective survey about real-life chatbot usage and experiences. We then introduce the model, review how the existing literature addresses the problem, and close by describing our experimental approach.

\subsection{Retrospective Survey}
To better understand key decision trade-offs faced by customers when choosing between chatbots and live agents, we conducted two waves of a retrospective survey on Prolific ($N=400$, see Appendix EC.1 for details).  Participants were asked to recall recent customer service interactions involving either a chatbot or a live agent, including details about wait times, issue resolution success, and overall satisfaction. The survey revealed significant differences in wait times to access customer service: about 77\%–79\% of chatbot users reported waits under one minute, compared to only 24\%–33\% for live agent interactions. However, chatbots resolved customer requests far less reliably, with success rates ranging between 34\%–42\%, compared to approximately 79\%–87\% for live agents. Overall satisfaction ratings were consistently higher for live agents (3.1 out of 5) relative to chatbots (2.2 out of 5). These results suggest the following. First, despite technological advancements, chatbot customer service experiences continue to be rated more poorly relative to live agent experiences. Understanding and mitigating adoption hurdles of chatbot technology thus continues to be an important practical concern for service managers. Second, the survey data highlight a key trade-off in customer channel choice: chatbots offer minimal wait times but fail frequently, while live agents require longer waits yet resolve requests more reliably. These insights guide both our model of channel choice and our experimental design.  

\subsection{Stylized Model of Channel Choice}
\begin{figure}[H]
\centering
\caption{Channel Choice} 
\label{fig:overview}
\includegraphics[width=0.85\textwidth]{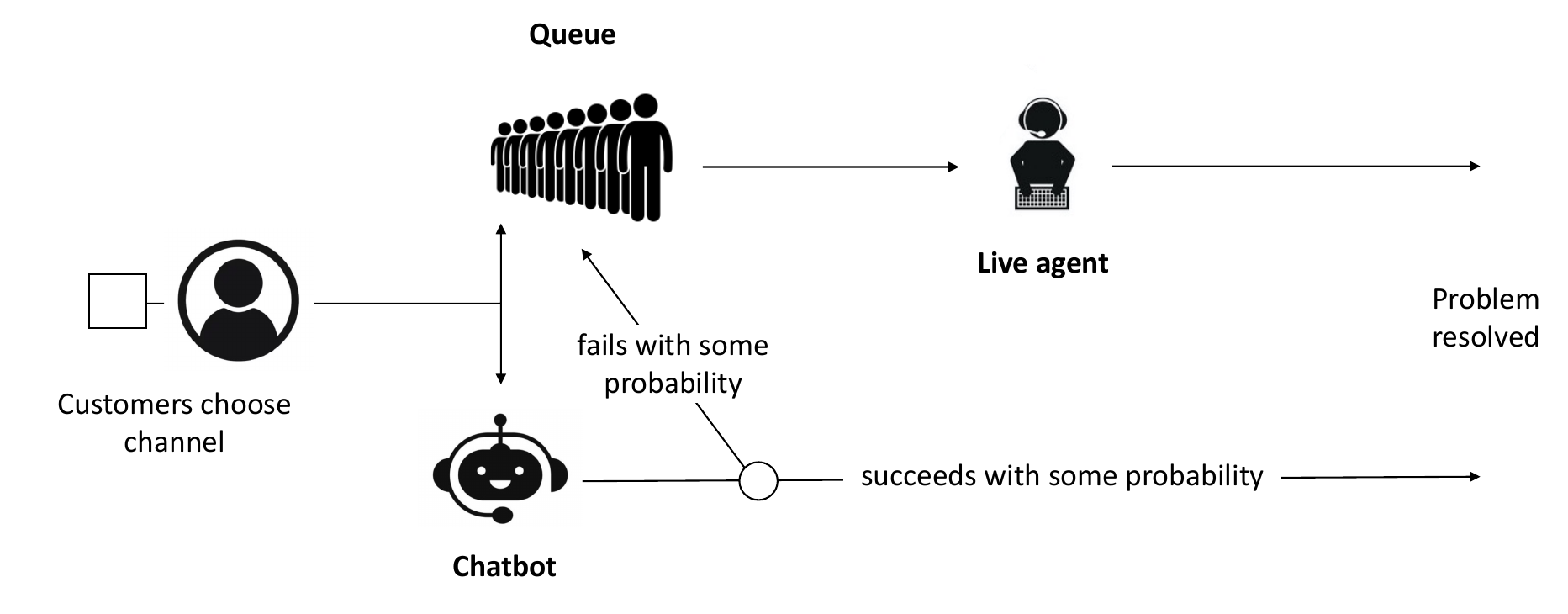}
\end{figure}

Building on our survey findings (\textsection 2.1), we model chatbot request resolution as a gatekeeper service process \citep{shumsky2003, freeman2017, hathaway2022}  -- a multi-stage process with an imperfect initial stage and a potential second stage with an expert service provider. Figure~\ref{fig:overview} illustrates the choice between a single-stage  live agent channel and a two-stage chatbot channel, where the chatbot acts as a gatekeeper. Consider first the live agent channel. The customer must wait in line, after which the server resolves the request with certainty, and the customer exits.\footnote{In practice, a small portion of requests handled by the live agent -- between 13\% and 21\% based on our survey results -- may be transferred to a second, expert agent. We do not examine such scenarios here to focus on the key trade-offs and to simplify choices for our experimental participants. However, such scenarios as well as other extensions of the decision problem in Figure \ref{fig:overview} are studied analytically in \cite{dada2025}.} Next, consider the chatbot channel. There is no queue, so the customer proceeds immediately to the chatbot interaction. Because the chatbot's problem-solving skills are limited, it only succeeds probabilistically. If the chatbot succeeds, the customer exits immediately. If it fails, then the customer waits in line before being served by a live agent, after which the request is resolved and the customer exits. Notably, the duration of the wait in line can be channel-dependent, and in our experiments and structural model we will consider priority queue designs that give chatbot customers a priority bump.%\footnote{We will further assume that the customer \textit{must} choose between the two channels, i.e., that the value of receiving customer service (net the waiting costs) is greater than the outside option of foregoing service or balking.}
\subsection{Related Literature}
Standard economic reasoning used in traditional queue joining models \citep[see][]{naor1969,allon2018} suggests that, when presented with a choice like the one in Figure \ref{fig:overview}, customers will select the channel that minimizes the expected time they would spend in the system. However, prior work in marketing, behavioral operations, and decision theory has identified several behaviors that suggest potential deviations from expected-time minimization in this setting. We discuss this work below.

\subsubsection*{Risk Aversion} First, the channels in Figure \ref{fig:overview} differ in the amount of risk they entail.  While risk preferences for money have been extensively studied \citep{holt2001,eckel2008,harrison2008}, relatively little is known about how individuals manage risk in the time domain. Some studies invoking Prospect Theory \citep{kahneman1979,tversky1992} suggest that because time expenditures are viewed as losses, individuals may be risk-neutral \citep{kroll2008} or even risk-seeking with respect to time \citep{abdellaoui2014}. However, more frequently, research finds risk-averse behavior in the time domain \citep{leclerc1995,festjens2015risk,flicker2022}. Importantly, much of this work relies on hypothetical decisions and examines longer time intervals than those typical in customer service settings. In contrast, our experiments are incentivized, so that participants' choices have consequences for how they spend time in the experiment. Moreover, because higher stakes have been shown to amplify risk aversion in financial decisions \citep{holt2001,harrison2008}, we use two treatment arms in our design: one with shorter and one with longer waiting time durations. This will allow us to explore how choices evolve as the stakes increase.

\subsubsection*{Transfer Aversion} %We use the term ``gatekeeper aversion'' to describe a preference for a more continuous, barrier-free service process with a single server, as opposed to a more fragmented process that may involve transfers and multiple service stages. 
Multi-stage waiting experiences were studied by \cite{carmon1996,kumar1997} and  \cite{kumar2021}. These papers show that customer satisfaction depends not only on the total time spent in the system but may fluctuate within and across waiting stages. Different from us, these studies focus on the affective response (self-reported in-process and ex-post satisfaction), while we follow the revealed-preference approach and study queue-joining behaviors. \cite{soman2003} show that the progression path to the goal (in our case, having one's service request resolved) can matter as much as the total time spent in the system. \cite{buell2021} shows that customers often exhibit last-place aversion in queues, suggesting that a focus on expected waiting times alone may oversimplify the waiting experience. \cite{althenayyan2022} look at fairness perceptions and show that customers experience in-queue delays differently depending on the source of the delay. Importantly, none of these studies look at multi-stage processes where the number of service stages is uncertain (i.e., gatekeeper processes), or where the nature of the server (human or algorithmic) is varied, leaving open the question of whether and how these factors interact. 

\subsubsection*{Algorithm Aversion} The algorithm aversion literature focuses on settings such as forecasting \citep{dietvorst2015,prahl2017,balakrishnan2022}, service delivery \citep{bastani2021,mejia2021,snyder2022}, order picking \citep{sun2022}, and recommendation settings \citep{yeomans2019}.  A common finding across many studies is that people exhibit aversion toward algorithms, particularly after seeing them err. However, some research also documents algorithm appreciation, where users prefer algorithmic advice, particularly for tasks perceived as objective and data-driven \citep{castelo2019, longoni2019}. Our approach differs from this literature in two ways. First, while the literature focuses on worker behavior, we focus on customer decisions; it is therefore not obvious whether the existing findings will hold in our setting.  Second, our survey evidence (\textsection 2.1) suggests a structural difference between channels: chatbot systems typically operate as gatekeeper systems, whereas live agent systems typically involve longer waits but only a single service stage. This is different from the standard experimental paradigm wherein algorithmic assistance is assumed to make the user better off \citep{prahl2017,yeomans2019}. Given these contextual differences, we develop a novel experimental approach %specifically designed 
to measure algorithm aversion in settings such as ours, where not only the nature of the server (human or algorithmic) but also the service design may differ across decision alternatives. In particular, we measure algorithm aversion by keeping the service design constant and by varying whether contextual information is presented to (or withheld from) participants.

\subsubsection*{Chatbot Adoption in Services}
Research specifically examining chatbot adoption in services remains relatively limited and often emphasizes anthropomorphism -- rendering the bot to be more human-like -- as a strategy to increase adoption \citep{sheehan2020, adam2021, schanke2021}.  Anthropomorphism can boost engagement and satisfaction by creating more natural conversational interactions, but its effectiveness varies across individual preferences and cultural contexts \citep{benke2022, luo2019}. More relevant for us is \cite{castelo2023}, who experimentally show that customers dislike service bots because they perceive them as a way for the service provider to cut costs at the customers' expense. Their study focuses on examining trust and fairness perceptions towards chatbots. In contrast, we explore operational dimensions of chatbot adoption, such as waiting times and chatbot effectiveness within a gatekeeper system, a common framework used in the service design literature \citep{shumsky2003,freeman2017,hathaway2022}. Our approach thus adds to the chatbot adoption literature by integrating the chatbot experience into an operational system, by studying factors that drive adoption decisions as well as by testing operational levers to increase adoption.

%\subsubsection*{Other Factors} There are other factors that may play a role in evaluating the cost of waiting. For example, waiting actively while being served may be preferred to waiting passively in line \citep{maister1984}. \cite{buell2011} show that customers have a preference for seeing that service is being performed while waiting. In our (customer service) setting, waiting episodes in which customers wait in line may generate a stronger aversive response than episodes in which customers are being actively served, even when the time spent is the same. Such preferences may favor the chatbot channel, which involves less waiting in line. 

\subsection{Experiment Design}
Our experiments serve three goals. First, they examine the extent to which the constructs identified in prior literature (risk aversion, transfer aversion, algorithm aversion, see \textsection 2.3) affect choices in the customer support setting, and explore potential ways in which these constructs interact (Experiment 1). Second, they test several managerial interventions to increase chatbot uptake (Experiment 2). Third, they allow us to evaluate different methodological approaches to eliciting and measuring algorithm aversion in the customer service setting (Experiment 3).

\noindent\textbf{All experiments}: In all experiments, participants make a series of binary choices between the live agent channel (Channel A) and the bot channel (Channel B), as shown in Figure \ref{fig:overview}. The choice is repeated for a range of problem parameters. Depending on the parametrization, Channel A or Channel B is the channel that minimizes total expected time spent in the system. Participants are incentivized to report their preferences truthfully by having to experience a subset of their choices (in real-time) before receiving their payments. Table \ref{tab:treat}  summarizes all three experiments, the hypotheses and the participant numbers.

\subsubsection*{Experiment 1:} This experiment examines channel choices in three settings: (1)  a contextualized setting in which the algorithmic nature of the chatbot is explicitly disclosed, (2) a neutral, context-free setting in which the live-agent/chatbot nature of the channel is not revealed, and (3) a context-free setting in which both channels present the decision-maker with a deterministic sequence of events, thus allowing us to isolate the role of risk preferences. More specifically, the experiment is organized as a between-subjects 3 (treatment) $\times$2 (time scale) design with the following treatments: 
\begin{enumerate}
\item[-]\textit{Context} Treatment. This treatment presents participants with a contextualized choice between two formats: the ``Live Agent'' and the ``Chatbot'' format. The sequence of stages in each channel is shown in Figure \ref{fig:overview}.  
\item[-]\textit{No Context} Treatment. This treatment is analogous to the \textit{Context} treatment but presents participants with a setting where the choice is between two unnamed waiting formats that are visually identical (but continue to differ in the sequence of waiting stages, as described in Figure \ref{fig:overview}).
\item[-]\textit{No Context, Deterministic} Treatment. This treatment is analogous to the \textit{No Context} treatment in that it does not present participants with any contextual information. However, different from the \textit{No Context} treatment, Channel B is now deterministic. This will allow us to separately identify the presence of risk aversion and transfer aversion. 
\end{enumerate}

As noted earlier, we use two between-subject treatment arms: a treatment arm with short time durations and a treatment arm with longer time durations. The size of the stakes (in our case, waiting time durations) is a commonly examined dimension in a variety of experiments studying both financial and time-relted choices \citep[See, for example,][]{holt2001,abdellaoui2014}; we therefore included two time scales in our design. In the short duration conditions, the average duration of a stage is 20 seconds across decisions, while in the long duration conditions it is 40 seconds. This results in average total waits of 40 seconds (resp.: 80 seconds) in the short (resp.: long) treatment arm. 

\subsubsection*{Experiment 2:}  In Experiment 2 we focus on the practical implementation challenges faced by firms deploying chatbots. We continue to use the \textit{Context} treatment as our baseline for comparisons. Against that baseline, we examine the effects of two manipulations: one related to how the chatbot capabilities are presented to the user, and a second one, related to how the waiting times are displayed in each channel:

\begin{enumerate}
\item[-]\textit{Context + No Transparency} treatment. In this treatment the chatbot always suggests a resolution to the issue,  regardless of whether it is viable. This is in contrast to the \textit{Context} treatment in which the chatbot is transparent about being able (or not) to resolve a given issue. 
\item[-]\textit{Context + Nudge} treatment. This treatment is analogous to the \textit{Context} treatment but adds the expected waiting times (in line + in service) for each channel. The expected waiting times serve as a nudge for the decision-maker to choose the channel that minimizes expected waiting duration.
\end{enumerate}

As before, we examine a treatment arm with short time durations and a treatment arm with longer durations. Most importantly, all decisions in the \textit{Context + No Transparency} and \textit{Context + Nudge} treatments are mathematically identical to the \textit{Context} treatment, with the only difference being how the choices and the interactions are presented.

 \begin{table}[bt!]
\caption{Experiment Overview}
\label{tab:treat}
\renewcommand{\arraystretch}{1.3}
\footnotesize 
\centering
\begin{tabular}{L{3.2cm} C{2.55cm} C{2.55cm} C{4.8cm} C{2.5cm}}
\toprule
\textbf{Objectives} & \multicolumn{2}{c}{\textbf{Treatments (Between-subject)}} & \textbf{Treatment Description} & \textbf{No. of Subjects}\newline(Recruited/ Passed comprehension screening/ Passed consistency checks) \\ 
\midrule
\addlinespace 
\multirow{6}{3cm}{\textbf{Experiment 1 (\textsection 3):}\newline What are key drivers of chatbot uptake in customer service?} 
& \uline{Short time durations:} & \uline{Long time durations:} & & \\
&  \textit{Context} &  \textit{Context} & Contextualized channel choice & 270/ 252/ 207 \\
&  \textit{No Context} &  \textit{No Context} & Context removed  & 238/ 227/ 183 \\
& \textit{No Context, Deterministic} & \textit{No Context, Deterministic} & Context and uncertainty removed & 263/ 253/ 207 \\
\addlinespace \addlinespace
\hdashline
\addlinespace
\multirow{5}{3cm}{\textbf{Experiment 2 (\textsection 4):} \newline What can firms do to increase chatbot uptake?} 
& \uline{Short time durations:} & \uline{Long time durations:} & & \\
%&  \textit{Context} & \textit{Context} & --- & --- \\
&  \textit{Context + No Transparency} &  \textit{Context + No Transparency} & Chatbot attempts all requests instead of admitting to not having a solution & 271/ 254/ 213 \\
&  \textit{Context + Nudge} &  \textit{Context + Nudge} & Added average waiting time information & 268/ 252/ 214 \\
\addlinespace \addlinespace
\hdashline
\addlinespace
\multirow{2}{3.2cm}{\textbf{Experiment 3 (\textsection 5):} How does the nature of the service process  affect algorithm aversion?} 
& \multicolumn{2}{c}{\uline{Short time durations:}} & & \\
%& \multicolumn{2}{c}{\textit{Context}} & --- & --- \\
& \multicolumn{2}{c}{\textit{Context + Live}} & Real-time chat with human agent & 116/ 106/ 91 \\
& \multicolumn{2}{c}{\textit{Context + Hold}} & Channel-specific interaction mode & 106/ 102/ 86 \\
\bottomrule
\end{tabular}
\end{table}

\subsubsection*{Experiment 3}
The purpose of the third experiment is to take a broader, more methodological view of characterizing the service process by examining alternative ways to measure algorithm aversion in a controlled experimental setting. To do so, we run two treatments that make the qualitative differences between the chatbot and live agent channels more salient:

\begin{enumerate}
\item[-] \textit{Context + Live} treatment. In this treatment the chatbot channel continues to use click-based prompts, while the live agent channel is staffed by a human research assistant.
\item[-] \textit{Context + Hold} treatment. In this treatment the chatbot channel continues to use click-based prompts, while participants in the live agent channel must hold down a button to complete the service process.
\end{enumerate}

Comparing chatbot uptake in these treatments to our baseline (\textit{Context} treatment) allows us to test whether our initial experimental findings are robust to more realistic implementations of service interactions and helps broaden our methodological contributions.

\section{Experiment 1: Adoption Hurdles}\label{sec:exp:1}
In Experiment 1 we focus on unpacking the drivers of the choice between a live agent service channel and an algorithmic chatbot. %To do so, we examine behaviors in three settings: one with a contextualized framing where participants choose among two service channels (\textit{Context} treatment), a second one where the same choices are made but the setting is void of context (\textit{No Context} treatment) and a third one in which the time spent in both channels is deterministic (\textit{No Context, Deterministic} treatment).

\subsection{Methodology}
%\subsubsection{Channel choice}  
%We begin by summarizing the channel choice shown in Figure \ref{fig:overview}. To access the server in Channel A, the participant first needs to wait in line.  The wait in line takes $t^{A}_{line_1}$ seconds, and the wait in service takes $t^{A}_{serve_1}$ seconds.  The server always succeeds in resolving the request and the participant exits the system. In Channel B, there is no line to start service, so that the participant immediately proceeds to interacting with the chatbot and spends $t^{B}_{serve_1}$ seconds to complete this interaction. The chatbot has limited problem-solving skills, resulting in some portion of chatbot interactions being redirected to a live agent, with $p^{B}$  denoting the probability of chatbot success.  If the chatbot succeeds in resolving the request,  the participants exits the system. If the chatbot fails,  the participant has to wait in line for $t^{B}_{line_2}$ seconds and then in service with a live agent agent for $t^{B}_{serve_2}$ seconds.  After that, the participant exits the system. All parameters are known to the participants' when they make their channel choices. 

\subsubsection{Participants, Pre-tests and Treatments}
A total of 771 participants were recruited on Prolific to participate in the experiment; 732 of them passed the screening questions, of which 597 passed the consistency checks. Additionally, 221 participants were recruited to participate in the pre-tests to the experiment, of which 213 passed all the checks. The pre-tests were designed to validate our \textit{Context} manipulation and to position our findings within the broader behavioral literature on algorithm aversion. All participants were US-based with an approval rating of at least 98\%, and were restricted to participating in one session only. All experiments were programmed in oTree \citep{chen2016}.  %We will base all our analyses on the sample of participants who passed both comprehension and consistency checks.%\footnote{The reported effects and significance levels are unchanged if we include participants who did not pass all consistency checks, but the standard errors are somewhat higher.%However, excluding inconsistent participants from the data is preferred practice in the experimental literature \citep[see][for comprehensive discussion]{charness2013}.} 

As noted earlier, the experiment consisted of three treatment conditions: \textit{Context}, \textit{No Context}, and \textit{No Context, Deterministic}. In addition, there were two treatment arms: one with short time durations and a second one with longer time durations (doubled times).  Each participant was randomly assigned to one of the treatment arms (short/long) and to one of the treatment conditions within that arm. %\footnote{To validate our measures, we also conducted a pre-test. A total of 114 Prolific workers  were recruited for this pre-test (113 after excluding one participant who did not pass comprehension checks). The results of this pre-test are discussed in \textsection 3.3.1.} 
Figure 2 shows the sequence of screens in each treatment and introduces notation for the relevant parameters. In all conditions, waiting in line ($t^A_{line},t^B_{line}$) was programmed to look the same. However, interactions between the participant and the server ($t^A_{serve},t^B_{serve_1},t^B_{serve_2}$) depended on the treatment. In the \textit{Context} treatment, choices were contextualized. In the instructions and on choice screens, participants were explicitly told that they were choosing between a live agent and a chatbot. Further, the interaction with each type of server consisted of channel-specific prompts (See EC for details). In this condition, we expect to see all three aversions: transfer  aversion, risk aversion and algorithm aversion. In the \textit{No Context} treatment conditions, choices were context-free and the interaction between participants and servers was programmed to look identical across channels (see Figure 2b). %The prompts appeared in equally spaced intervals with a total of three prompts per interaction, and the progress bar resumed only after the participant responded to the prompt. 
In this treatment, we expect to see transfer aversion and risk aversion, but no algorithm aversion.  Finally, in the \textit{No Context, Deterministic} treatment, the participant always experienced both service stages in Channel B whereas in the other two, the participant experiences one service stage with probability $p^B$ and two with probability $1 - p^B$. However, the durations of the service stages were adjusted such that the expected times in each decision were identical to the corresponding decision in the non-deterministic treatments. Given that both context and uncertainty is removed, in this treatment we expect to see transfer aversion only. %We next develop our experimental hypotheses.

\begin{figure}[H]
\caption{Flow of Waiting Stages by Treatment}
\label{fig:flow:new}
\centering  
\vspace{0.1cm}  

\begin{subfigure}[b]{0.8\textwidth}
    \caption{\small\baselineskip=9pt \textit{Context} Treatment}  
    \includegraphics[width=\textwidth]{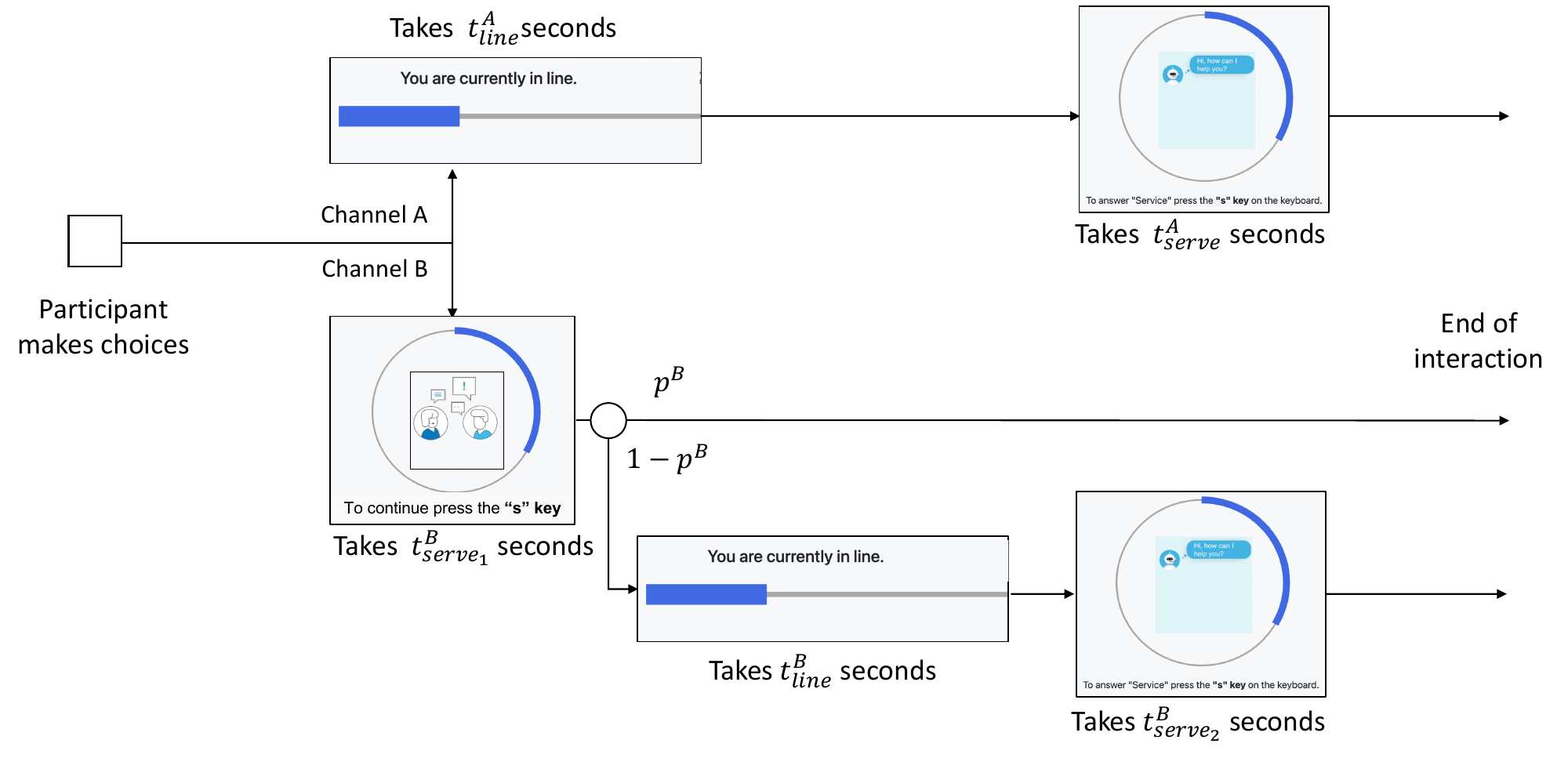}
\end{subfigure}
\vspace{0.8cm}

\begin{subfigure}[b]{0.85\textwidth}
    \caption{\small\baselineskip=9pt {\textit{No Context} Treatment}}
    \includegraphics[width=\textwidth]{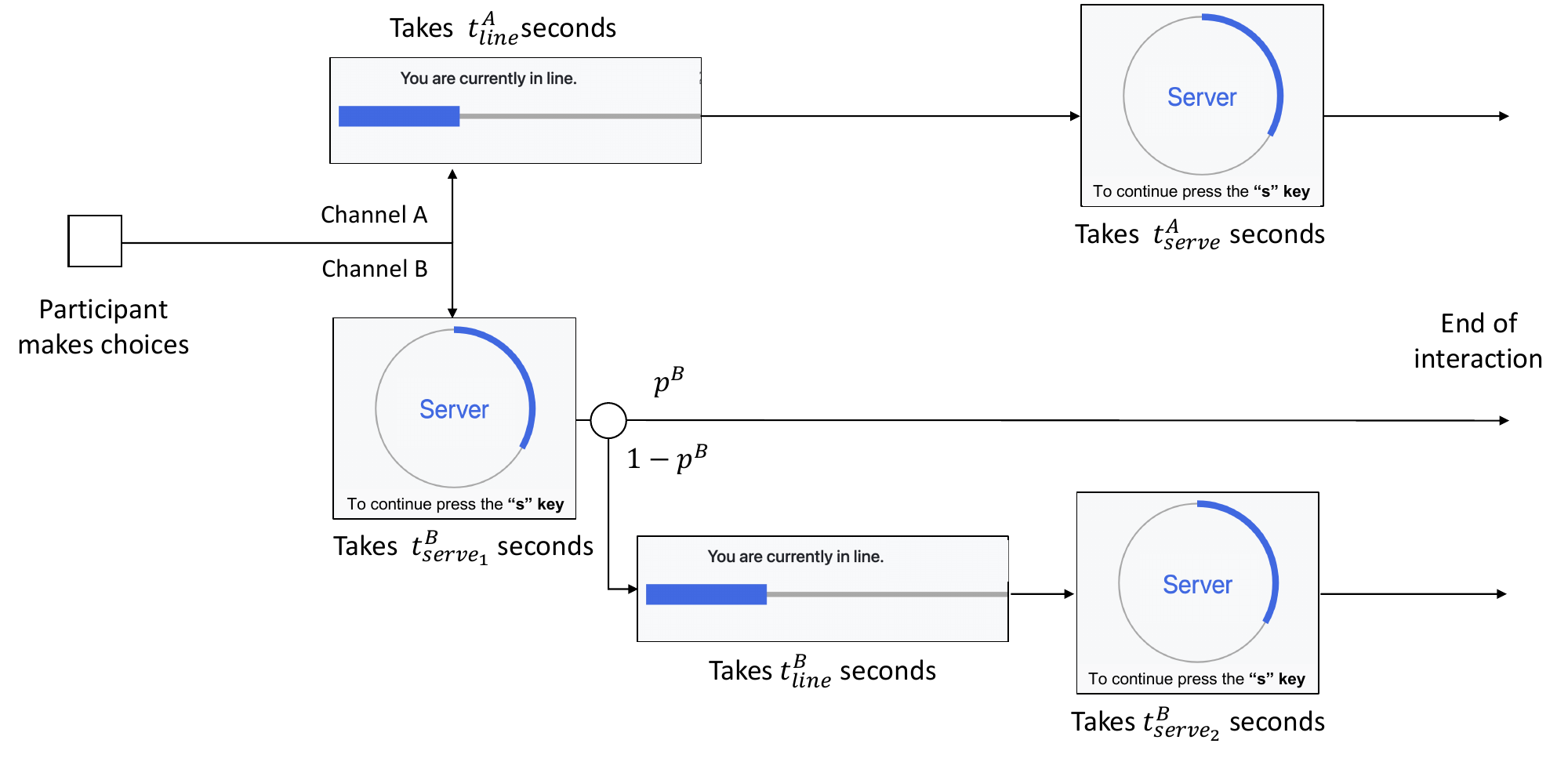}
\end{subfigure}
\vspace{0.8cm}

\begin{subfigure}[b]{0.85\textwidth}
    \caption{\small\baselineskip=9pt {\textit{No Context, Deterministic}  Treatment}}
    \includegraphics[width=\textwidth]{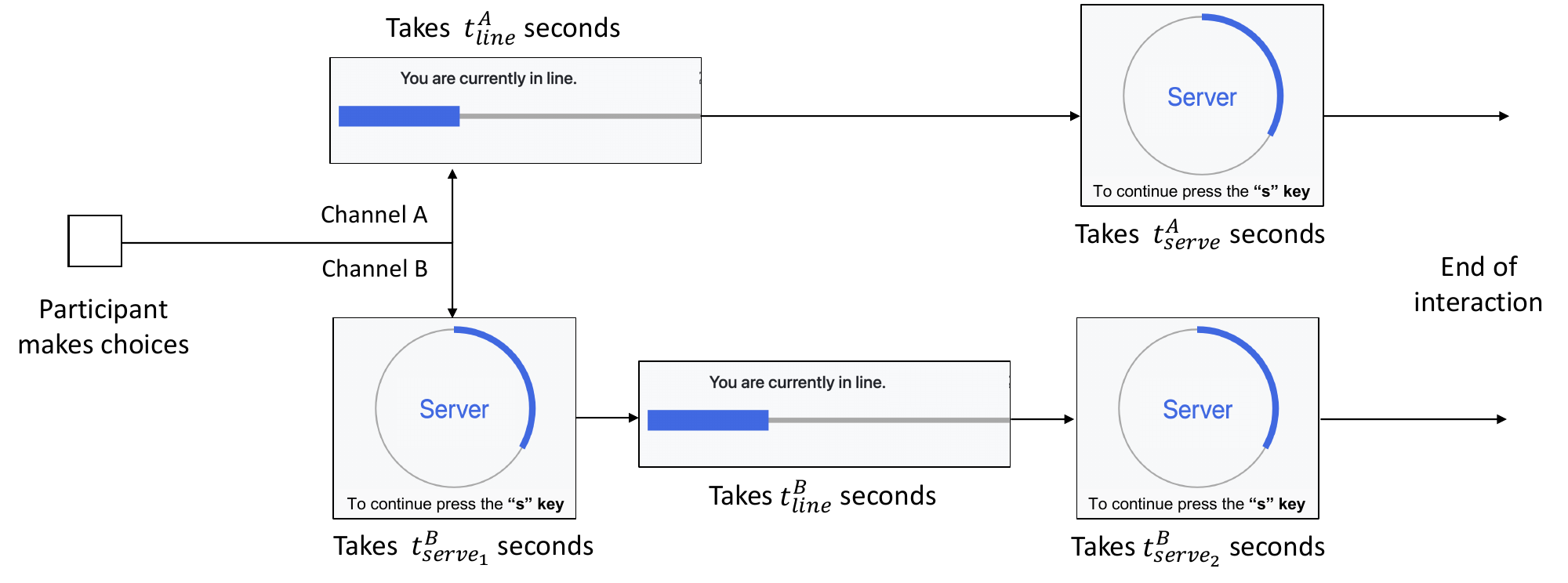}
\end{subfigure}
\end{figure}

\subsubsection{Instructions and Demo}  
Figure 3 describes the experiment protocol. After being randomly assigned to a treatment, reading the instructions and answering comprehension questions, participants experience a demo of both channels. The Channel A demo is parametrized with $t^{A}_{line} = t^{A}_{serve} = 20$ seconds for short scale treatments and 40 seconds for long scale treatments. The Channel B demo follows and includes both the successful and failed chatbot resolution scenarios, with  $t^{B}_{serve_1} = t^{B}_{line} = t^{B}_{serve_2} = 20$ seconds for short (40 seconds for long). The demo is visually representative of the actual experience in each channel and thus differs by treatment. Screenshots of the interface are in Appendix EC.3.

\subsubsection{Decisions and Parameters}  
After the demo, participants make a total of 33 decisions, subdivided into three decision sets of 11 decisions. The sequence of decision sets was randomized to control for any order effects. Each decision is a binary choice between Channel A and Channel B. The chatbot capability $p^{B}$, the chatbot service time $t^{B}_{serve_1}$, and the waiting time after potential chatbot failure $t^{B}_{line}$ differ across the 33 decisions. In particular,  $p^{B}$ ranges from 0.25 to 0.75 in increments of 0.05. In short duration conditions, $t^{B}_{serve_1}$ ranges from 10 to 30 seconds in 2-second increments, and $t^{B}_{line}$ ranges from 0 to 40 seconds in 4-second increments. Long duration conditions double these time parameters: $t^{B}_{serve_1}$ ranges from 20 to 60 seconds, and $t^{B}_{line}$ from 0 to 80 seconds. Within each decision set, Channel A minimizes expected waiting times in the first five decisions, Channel B minimizes expected waiting times in the last five decisions, and both channels yield identical expected waiting times in the sixth decision. Complete parameter listings for all 33 decisions are in the Electronic Companion. 

\begin{figure}[t!]
\centering
\caption{Experiment Protocol}
\label{fig:design}
\includegraphics[width=\textwidth]{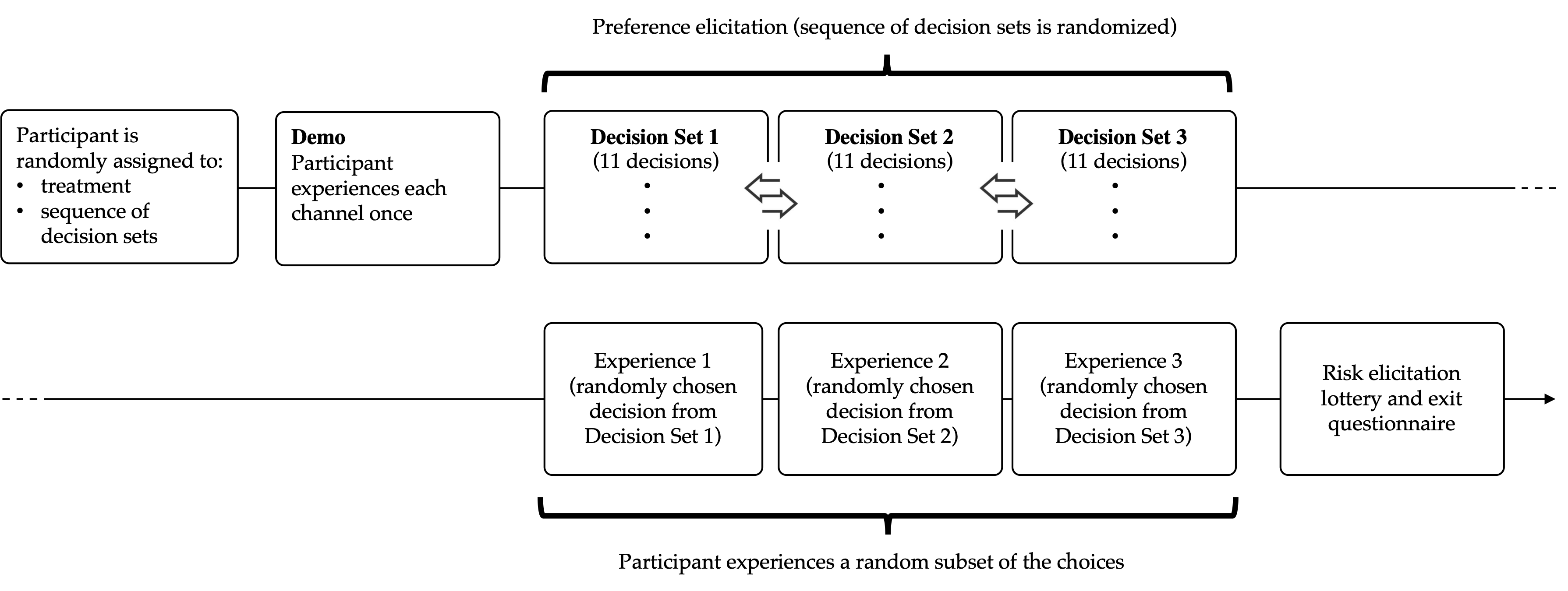}
\end{figure}
\subsubsection{Elicitation}
Within each decision set, we used the Multiple Price Lottery mechanism \citep{holt2001} to elicit preferences. The basic idea of this mechanism is to present participants with a list of binary decisions, where one of the alternatives becomes more desirable as one goes down the list. For example, in Decision Set 1 we varied the success rate of the chatbot ($p^{B}$)  in steps of 0.05 from 0.25 to 0.75, while all other parameters were held constant. Across all three decision sets, we kept constant the difference in expected times between the two alternatives for each decision within a decision set (i.e., Decision 1 in Decision Set 1 has the same expected time difference between channels as Decision 1 in Decision Set 2 and Decision 1 in Decision Set 3, and similarly for the remaining 10 decisions). 

\subsubsection{Incentives}
After a participant completed all their decisions, a subset of the participant's decisions was selected to be experienced in real time.  Specifically, one decision from each decision set was selected at random for the real experience, resulting in each participant experiencing three of their 33 choices prior to receiving their (fixed) dollar payment and exiting the experiment. Thus, participants were incentivized to report their true preferences. The average time spent in the experiment was 18 (resp.: 23) minutes in the short (resp.: long) time conditions. The average payment was \$6.65 (resp.: \$8.14).\footnote{In addition to the main task, for which participants received a fixed participation payment, at the end of the experiment we elicited the participants' risk aversion (with respect to money) using an incentivized version of the Eckel-Grossman single lottery test \citep{eckel2002,eckel2008}, which could earn participants up to an additional \$2.}

\subsection{Hypotheses}
Our hypothesis testing approach is summarized in Table \ref{tab:hyp}. Before testing our hypotheses we will perform a series of pre-tests in which Channels A and B will have identical performance and sequence of stages and will only differ in the visual cues. The pre-tests will help identify algorithm aversion in the absence of process or performance-related differences. Subsequently, we will test for the presence of each type of aversion by examining the differences in mean Channel B uptake across treatments.

\begin{table}[h]
  \centering
  \footnotesize
  \centering
    \caption{Experiment 1: Hypothesis Testing Approach}
    \label{tab:hyp}
        \begin{tabular}{lcccc}
      \toprule
       &  & \multicolumn{3}{c}{\textbf{Treatment}} \\
        \cmidrule(lr){3-5}
      & \textbf{Pre-tests} & \textit{No Context, Deterministic} & \textit{No Context} & \textit{Context}  \\
       \cmidrule(lr){2-2} \cmidrule(lr){3-5}
        &   & Transfer aversion (H1.1) & Transfer aversion  & Transfer aversion \\
        &  &  & + Risk aversion (H1.2) & + Risk aversion \\
        & Algorithm aversion  &   &  & + Algorithm aversion (H1.3)  \\
      \bottomrule
    \end{tabular}
    \end{table}
    
Consider first the \textit{No Context, Deterministic} treatment. In this treatment channels A and B are visually identical, and there is no risk in either channel. Therefore, transfer aversion is the sole mechanism that may drive potential deviation from expected-time minimization. While the literature on multi-stage waiting experiences is limited (see \textsection 2.3), some work suggests that departures from a single-stage service process can lower satisfaction (e.g., \citealt{soman2003,kumar2021}).  Therefore, we hypothesize:

\vspace{0.3cm}

\noindent\textbf{H1.1 (Transfer Aversion):} Average Channel B uptake in the \textit{No Context, Deterministic} treatment is below 5.5 (risk-neutral theory prediction).\footnote{Recall that there are 11 decisions in each decision set. Given the parameterizations in the experiment, expected time minimization predicts that a decision-maker switches from Channel A to Channel B in decision 6 where both channels offer the same. If we assume random tie breaks, 5.5 is the theory prediction. However, our results are robust to alternative tie-breaking procedures for the sixth decision, i.e., to using 5 out of 11 or 6 out of 11 as our theory benchmark. See Electronic Companion for full listings of parameters in all treatments.}

\vspace{0.3cm}

\noindent Recall that in the \textit{No Context} treatment Channel B entails risk, while in the \textit{No Context, Deterministic} treatment, neither channel is risky. Further, in both of these treatments, channels A and B are visually identical and include no contextual cues. Therefore, a treatment comparison between the \textit{No Context} and the \textit{No Context, Deterministic} treatments will isolate risk aversion.

\vspace{0.3cm}

\noindent\textbf{H1.2 (Risk Aversion):} Conditional on expected waiting times in each channel,  participants in the \textit{No Context} treatment choose Channel B at a lower rate  than in the \textit{No Context, Deterministic} treatment.

\vspace{0.3cm}

\noindent  Our \textit{No Context} and \textit{Context} treatments differ only in how the channels are presented, i.e., whether the algorithmic nature of Channel B is made salient. Therefore, a comparison of these two treatments will allow us to isolate algorithmic aversion. 
\vspace{0.3cm}

\noindent\textbf{H1.3 (Algorithm Aversion):} Conditional on expected waiting times in each channel,  participants in the \textit{Context} treatment choose Channel B  at a lower rate  than in the \textit{No Context} treatment.
 
\vspace{0.3cm}

To test H1.1 we perform one sample $t-$tests, comparing empirically observed chatbot uptake against 5.5, the theoretically optimal uptake under expected-time minimization. To test H1.2-H1.3, we perform random effects logit regressions and examine the significance of treatment coefficients. Because our hypotheses make a directional prediction, we will report one-sided $p-$values adjusted for multiple hypothesis testing using the Bonferroni-Holm procedure for each family of hypotheses \citep{holm1979,list2019}.

\subsection{Results}
We begin by reporting the results of two pre-tests designed to validate our \textit{Context} manipulation and to better position our findings within the broader behavioral literature on algorithm aversion. After reporting the results of these pre-tests, we present descriptive statistics and formal hypothesis tests.

\subsubsection{Pre-tests}
The first pre-test ($N=113$) examined whether participants had any intrinsic preference for (or bias against) the visual stimuli used to represent the human or the algorithmic channel. Participants experienced two interactions: a 20-second interaction with a chatbot and a 20-second interaction with a human agent, presented in randomized order. (See Figures 2b and c for theses visual stimuli, which are also used in the \textit{Context} treatments.) Participants then selected one provider (chatbot or live agent) for an additional 40-second interaction; this choice served as our outcome variable. A total of 56 of 113 participants chose the live agent (49.56\%), which is not significantly different than 50\% (Proportion test, $p=0.904$). Thus, neither set of visual stimuli produced a bias towards one channel over the other.

The second pre-test ($N=100$) was similar in that each participant made a single decision after experiencing each channel once. However, now participants chose between two identical \textit{gatekeeper} processes, each offering a 50\% chance of immediate resolution following an initial 40-second interaction, and a 50\% chance of requiring an additional 40-second wait in line plus a 40-second interaction with a second-stage human agent. The only difference between the two channels was the labeling and visuals of the initial server as either live agent or chatbot. Results again revealed no significant preference for either channel: 49 out of 100 participants chose the live agent (49\%), vs. 51 participants (51\%) chose the chatbot (Proportion test, $p=0.920$). Thus, our experimental stimuli do not produce a detectable bias towards humans or algorithms when performance (error rates) and processes are held constant between human and algorithmic alternatives. However, as will become clear later in this section, algorithm aversion can still play an important role as an \textit{amplifier}, reinforcing gatekeeper aversion and further reducing chatbot adoption when the chatbot is associated with existing process and performance differences.

\subsubsection{Descriptive Statistics}
Figure 4 shows the share of participants choosing Channel B (representing the chatbot) in each of the six conditions, conditional on the expected time difference between the two channels. Within each decision set, Channel B becomes increasingly more attractive for higher-numbered decisions, with Decision 6 marking the indifference point. Several observations are in order. First, in all six conditions, chatbot uptake is substantially below the expected-time minimization benchmark. This is particularly visible in the right half of each graph (where time savings in Channel B are positive), with a gap ranging between 10 and 70 percentage points. Further, much of this behavior appears to be tied to the process-related features of the choice, with a substantial Channel B underutilization even in the absence of context. Indeed, adding context further reduces chatbot uptake, but it does so only by a small margin (0 to 10 percentage points), with a more pronounced change in panel b. These observations offer some preliminary support for hypotheses H1.1-H1.2 but suggest that H1.3, i.e., algorithm aversion, may only be observed for the longer duration conditions, i.e., when the stakes are higher. Finally, comparing panel a and panel b, longer time durations decrease chatbot uptake in all three treatments (with the difference being particularly large for the \textit{No Context} and \textit{Context} conditions). While not part of our formal hypothesis development, we will nonetheless test the significance of this difference in post-hoc analysis (\textsection 3.3.4).

\begin{figure}[b!]
\centering
\caption{Channel B (Chatbot Channel) Uptake in Experiment 1}
\label{fig:bars_s1}
\includegraphics[width=0.99\textwidth]{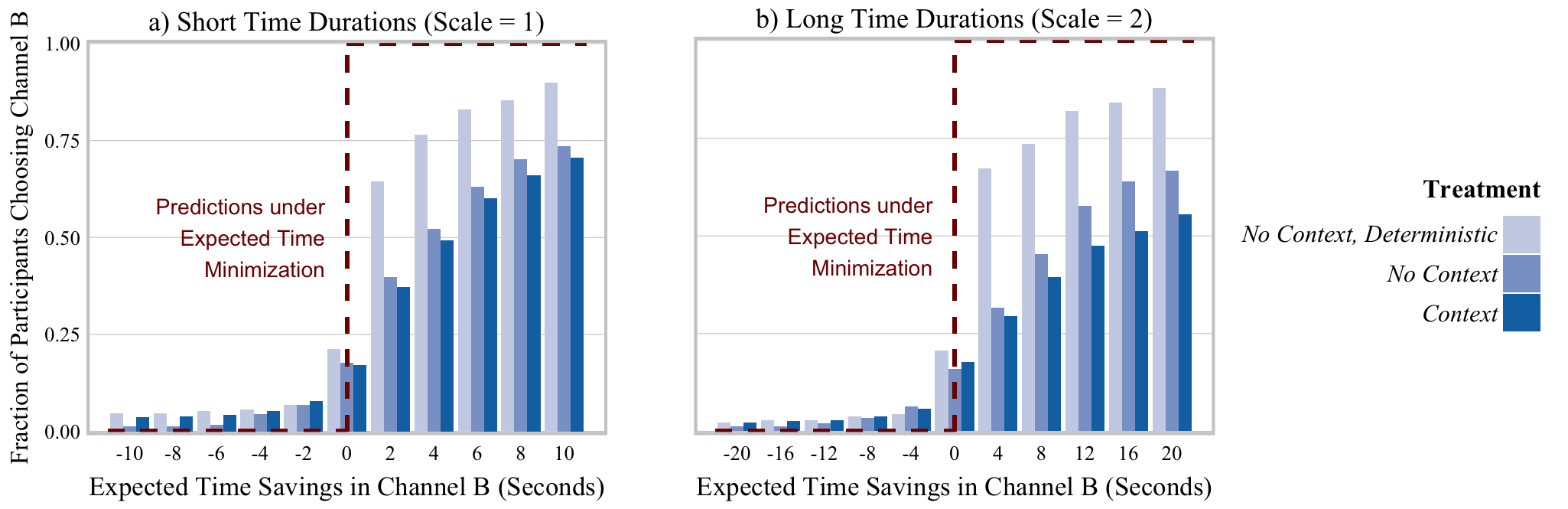}
\end{figure}

\subsubsection{Hypothesis Tests}
We next test H1.1 - H1.3, i.e., examine whether chatbot uptake is below what expected-time minimization would predict, and whether it is affected by the presence of context. Because we are simultaneously testing three hypotheses, it is appropriate to adjust the reported significance levels for multiple hypothesis testing.  

We begin by testing H1.1, i.e., the presence of transfer aversion. To test H1.1 we use $t-$tests and compare observed channel uptake with 5.5 (out of 11), the expected-time minimization benchmark. Our data strongly reject H1.1: average Channel A uptake is  4.47 (resp.: 4.32) in the short (resp.: long) time durations condition; both values are significantly smaller than 5.5, $p\ll0.001$. To test H1.2 we use random effects logit regressions. The regression coefficients are reported in Table \ref{tab:reg:exp1}. Column 1 presents the full data set. We use the  \textit{No Context} treatment as the omitted variable, because this treatment is used for comparisons in both H1.2 and H1.3. Note first that the difference between the \textit{No Context} and \textit{No Context, Deterministic} treatments is statistically significant ($p\ll0.001$). The absence of uncertainty significantly increases Channel B uptake. However, the difference between the \textit{Context} and \textit{No Context} treatments is not statistically significant ($p=0.179$). Columns 2-3 replicate the analysis, but focus on either the short durations conditions (col. 2) or the long durations conditions (col. 3). The same pattern of results emerges, with the difference being that the effect sizes are somewhat higher for the long time durations. Indeed, the \textit{Context} treatment dummy is significant at $p=0.046$ in col. 3, suggesting that algorithm aversion is present for longer time durations. 

\noindent \textit{\textbf{Result 1:} H1.1-H1.2 are supported. We find evidence for both transfer aversion and risk aversion. H1.3 is partially supported. Adding customer service context significantly reduces chatbot uptake for long, but not for short time durations.}

\begin{table}[b!]
\renewcommand{\arraystretch}{1.1}
\caption{Channel Preferences in Experiment 1}
\label{tab:reg:exp1}
\footnotesize
\centering
\begin{tabular}{lC{2.2cm}C{2.2cm}C{2.2cm}}
\toprule
& (1) & (2) & (3) \\
\multicolumn{1}{r}{Dependent Variable:} & Channel B (Chatbot Channel)& Channel B (Chatbot Channel)& Channel B (Chatbot Channel)\\
\midrule
\textit{No Context}  Treatment & Omitted category & Omitted category & Omitted category \\
\addlinespace
\textit{No Context, Deterministic} Treatment & 2.134*** & 1.800*** & 2.585*** \\
& (0.401) & (0.522) & (0.668) \\
\addlinespace
\textit{Context} Treatment & -0.560 & -0.054 & -1.116** \\
& (0.416) & (0.545) & (0.661) \\
\addlinespace
\textit{Time scale $= 2$} & -0.821** & & \\
& (0.323) & & \\
%\addlinespace
%Constant & -7.253*** & -6.252*** & -9.760*** \\
%& (0.822) & (0.977) & (1.375) \\
\midrule
Channel Performance Controls & Yes & Yes & Yes \\
Demographic Controls & Yes & Yes & Yes \\
Sample & Full Sample & Time Scale $ = 1$ & Time Scale $ = 2$ \\
Observations & 19701 & 9504 & 10197 \\
Subjects & 597 & 288 & 309 \\
\bottomrule
\end{tabular}
\vspace{0.3em}

\begin{minipage}{0.8\textwidth}
\setlength{\baselineskip}{0.1\baselineskip}
\footnotesize \textit{Notes:} Random effects logit regression coefficients are reported. Dependent variable is channel choice (Channel B $= 1$). Standard errors are clustered at subject level. Decision set number and decision number within the decision set are controlled for. The following demographic variables are controlled for: age, gender, number of quiz errors and the Eckel-Grossman risk aversion measure (administered after the main task). H1 tests are one-sided, with $p-$values adjusted for multiple hypothesis testing using Bonferroni-Holm procedure. The remaining tests are two-sided. *** $p<0.01$ ** $p<0.05$ * $p<0.1$.
\end{minipage}
\end{table}

\subsubsection{Additional Analysis}
Most lab experiments, and ours, involve only a small number of relatively short interactions, which can limit external validity. To address this concern, we performed three sets of supplementary analyses. First, we check for potential learning effects. To do so, we verify that the decision set variable is not statistically significant. Indeed, based on regression specifications in Table \ref{tab:reg:exp1}, we find no evidence of learning, with all decision set coefficients being well above the significance threshold ($p>0.256$).  Second, we examine potential time scale effects. In particular, Figure 4 suggested that all three aversions increase with longer time durations. Column (1) in Table \ref{tab:reg:exp1} confirms that this effect is significant: the scale effect is negative and statistically significant at $p=0.011$, with chatbot adoption decreasing by an average 4.4 percentage points when durations are long. Thus, the duration of the interactions is an additional contributor to the willingness (or reluctance) to engage with chatbot technology. Finally, we compare participant behavior in the experiment with their responses to the post-experimental questions regarding everyday chatbot use. We find that participants who reported prior use of chatbots outside of the experiment showed significantly higher Channel B uptake in the experiment, with an increase between 13\% and 26\% relative to those participants who have had no prior experience with chatbots (rank sum tests $p<0.01$).

\subsection{Discussion}
The results of Experiment 1 offer several interesting insights. As expected, participants respond positively to improved chatbot performance and choose Channel B more frequently as its time (in line and in service) becomes shorter and as the chatbot's probability of success increases. At the same time, participants' choices deviate significantly from expected-time minimization: the majority of participants frequently chose the channel with longer expected waiting times, suggesting a significant aversion to gatekeeper systems both with and without uncertainty and both in the presence and in the absence of contextual cues.

In our pre-test (\textsection 3.3.1) we saw that context alone did not significantly drive channel preferences. Similarly, context had only minimal effects in our main experiment, when durations were short. However, when we increased the stakes (doubling the time durations in both channels), we observed significant algorithm aversion in addition to gatekeeper aversion. Thus, algorithm aversion can amplify peoples' reluctance to use a gatekeeper channel. From a theory standpoint, this suggests that algorithmic attitudes are malleable and can interact with structural biases around the gatekeeper process itself. This is different from the classic result that people are more sensitive to algorithmic than to human errors even when humans and algorithms perform equally well \citep{dietvorst2015}. If chatbots performed at the same level as live agents, our results suggest that their uptake will be closer to rational theory predictions. However, as long as chatbots remain limited in their capabilities (as is currently the case in practice; see \textsection 2.1), customers will continue to underutilize them -- both due to their aversion to gatekeeper processes and because chatbots have become emblematic of that very experience.

From a more practical standpoint, the result that algorithmic and process-related hurdles can be mutually reinforcing suggests that firms need to consider chatbots as part of a larger, multi-channel service design problem instead of treating chatbots as a standalone AI issue.  Building on this idea, in the next section we test two practical remedies aimed at increasing chatbot uptake.

\section{Experiment 2: Remedies}\label{sec:exp3}
We have so far seen that the bulk of Channel B (chatbot) underutilization is tied to the gatekeeper structure of the chatbot channel.  We have also seen that chatbot uptake may be further reduced for request types that involve longer durations. In this section, we propose and test two managerial levers to counteract these adoption hurdles. 

\subsection{Treatments}
Experiment 2 introduces two new treatments. As in Experiment 1, each treatment consists of two between-subject conditions: one with short and one with long (doubled) time durations. We continue to use the \textit{Context} treatment from Experiment 1 as our baseline for comparisons. The new treatments are the \textit{Context + No Transparency} and the \textit{Context + Nudge} treatment. In the \textit{Context + No Transparency} treatment the chatbot always suggests a solution to the request, regardless of whether it is able to resolve it successfully. This is in contrast to the \textit{Context} treatment, in which the chatbot is transparent about its capabilities and simply reports not being able to resolve an issue if this happens to be the case. See Electronic Companion for the relevant screens displayed in each treatment. In the \textit{Context + Nudge} treatment we provide participants with expected waiting times for each channel and decision. The parameters as well as the remaining prompts and instructions are unchanged relative to the \textit{Context} treatment. We recruited 539 participants for the new treatments via Prolific, of whom 427 participants passed all comprehension checks and consistency checks.
 
\subsection{Theory and Hypotheses}
To develop hypotheses we leverage the rich behavioral literatures in economics and operations. The \textit{Context + No Transparency} treatment is inspired by prior work on operational transparency, which has consistently shown that revealing the processes underlying service delivery increases trust and perceived value \citep{buell2011,buell2017}. \cite{balakrishnan2022} show that feature transparency reduces algorithm aversion in forecasting tasks. Somewhat different from this literature, we focus on outcome (as opposed to process) transparency. We expect that being transparent about the ability of the chatbot to resolve a given request may increase trust and thus increase chatbot uptake. Specifically, we compare a transparent chatbot that explicitly communicates its capabilities and limitations versus one that attempts to handle all requests without such disclosure. Formally, we hypothesize the following:
\smallskip

\noindent\textbf{H2.1 (Transparency):} Conditional on expected times in both channels, participants in the \textit{Context + No Transparency} treatment choose Channel B at a lower rate than in the \textit{Context} treatment.
\smallskip

Our second intervention (\textit{Context + Nudge}) leverages insights from behavioral economics about how people process complex choices under uncertainty. In particular, decision-makers often fail to optimally integrate outcomes and probabilities, instead focusing on particularly salient aspects of the choice \citep{arieli2011,aimone2016s,aimone2016}. The \textit{Context + Nudge} intervention is aimed at directing decision-makers' attention towards objective performance metrics (expected times)  and away from format preferences or channel biases. This intervention thus offers a lightweight ``nudge'' that could increase the share of people choosing the chatbot option, when this option helps them save time (in expectation). Formally, we test the following hypothesis:

\smallskip

\noindent\textbf{H2.2 (Nudge):} Conditional on expected times in both channels, participants in the \textit{Context} treatment choose Channel B at a lower rate  than in the \textit{Context + Nudge} treatment.

\subsection{Results}
As before, we present average chatbot uptake by treatment, and then test our hypotheses using random effects logit regressions.

\subsubsection{Descriptive Statistics}
Figure 5 shows chatbot uptake in each of the six conditions and in each of the 11 decisions. First, relative to the \textit{Context + No Transparency} treatment, transparency (\textit{Context} treatment) appears to increase chatbot uptake for short time durations (panel a), with increases between five and ten percentage points in the right half of the graph. This is consistent with H2.1. However, transparency appears to have only a minimal effect in the long time durations conditions. Second, consistent with H2.2, the nudge appears to increase chatbot uptake in both treatments, though the increase is quite small under short time durations (at most five percentage points) and quite strong under longer time durations (up to 20 percentage points). Further, note that the scale effect (reduced chatbot uptake for longer time durations) observed in Experiment 1, while still present on average, does not hold across all treatments. In particular, comparing the dark blue bars in panel a) with those in panel b), we observe that there is no discernible scale effect in the \textit{Context + Nudge} condition. This suggests that the effect of the nudge is quite dominant in focusing participant attention on the decision-theoretic fundamentals of the choice, bringing their decisions closer to the theoretic predictions. 

\begin{figure}[bh!]
\centering
\caption{Channel B (Chatbot Channel) Uptake in Experiment 2}
\label{fig:bars_s2}
\includegraphics[width=0.99\textwidth]{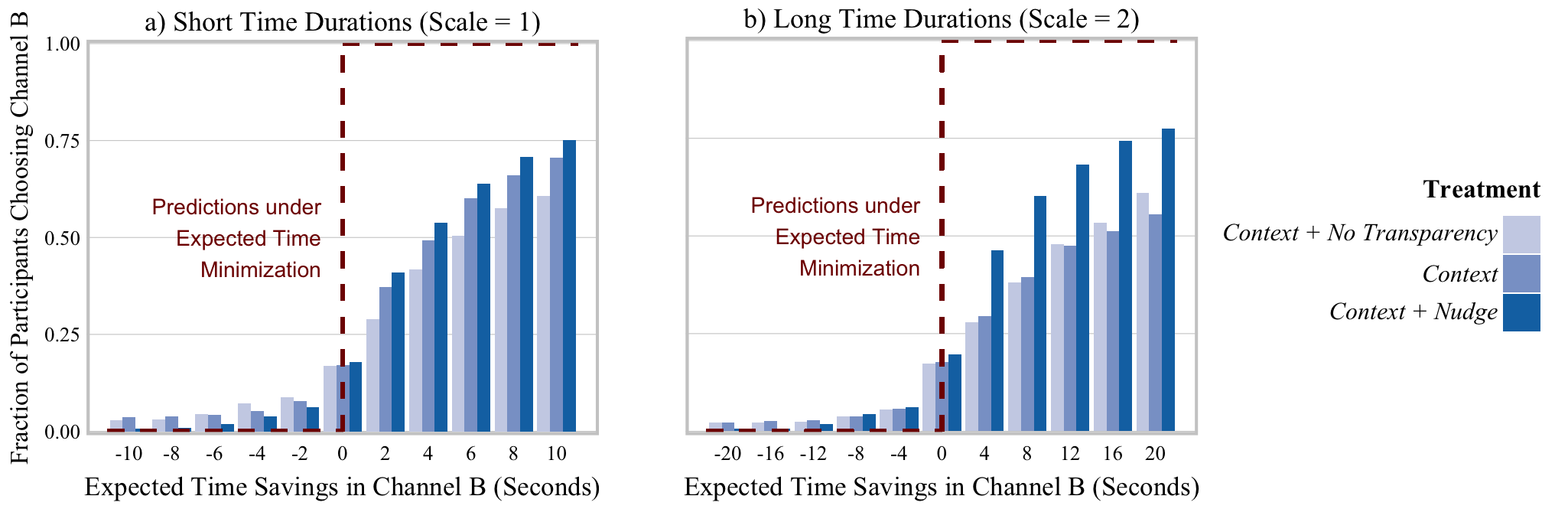}
\end{figure}
\subsubsection{Hypothesis Tests}
To test H2, we regressed channel choice on the treatment dummies. Table \ref{tab:reg:exp2} presents the estimates. As before, in Table \ref{tab:reg:exp2}  we report one-sided, Bonferroni-Holm adjusted $p-$values. In column (1) we present test results for the pooled sample. \textit{Context} treatment is used as the  baseline (omitted category). Several observations are in order. First, transparency appears to have little effect on chatbot uptake in the aggregate: as predicted, the \textit{Context + No Transparency} treatment coefficient is negative but the effect is not statistically significant ($p=0.138$). Second, the effect of the nudge is positive and statistically significant ($p=0.008$). In column (2) we focus on short time durations. Here, the effects of transparency are statistically significant: being transparent about chatbot capabilities significantly increases chatbot uptake relative to the condition where the chatbot always produces a solution ($p=0.019$). Further, the effect of the nudge is not statistically significant ($p=0.669$). Finally, column (3) focuses on the long durations conditions and shows that the effect of transparency is not statistically significant ($p=0.664$), while the nudge significantly increases uptake ($p\ll0.001$). %We next discuss potential reasons for these results. 
 
\smallskip
\noindent \textit{\textbf{Result 2:} H2.1 is partially supported. Chatbot channel uptake is increased under operational transparency, but the effect is only observed when time durations are short. H2.2 is supported. The nudge helps increase chatbot uptake.}

\begin{table}[bth!]
\renewcommand{\arraystretch}{1.1}
\caption{Channel Preferences in Experiment 2}
\label{tab:reg:exp2}
\footnotesize
\centering
\begin{tabular}{lC{2.2cm}C{2.2cm}C{2.2cm}}
\toprule
& (1) & (2) & (3) \\
\multicolumn{1}{r}{Dependent Variable:} & Channel B (Chatbot Channel) & Channel B (Chatbot Channel)& Channel B (Chatbot Channel)\\
\midrule
\textit{Context} Treatment & Omitted category & Omitted category & Omitted category \\
\addlinespace
\textit{Context + No Transparency} Treatment & -0.412 & -1.078** & 0.225 \\
& (0.378) & (0.521) & (0.529) \\
\addlinespace
\textit{Context + Nudge} Treatment & 0.883*** & -0.221 & 2.049*** \\
& (0.364) & (0.506) & (0.523) \\
\addlinespace
\textit{Time scale $= 2$} & -0.463 & & \\
& (0.302) & & \\
%\addlinespace
%Constant & -5.626*** & -2.951** & -8.448*** \\
%& (0.880) & (1.243) & (1.199) \\
\midrule
Channel Performance Controls & Yes & Yes & Yes \\
Demographic Controls & Yes & Yes & Yes \\
Sample & Full Sample & Time Scale $ = 1$ & Time Scale $ = 2$ \\
Observations & 20922 & 10428 & 10494 \\
Subjects & 634 & 316 & 318 \\
\bottomrule
\end{tabular}

\vspace{0.3em}
\begin{minipage}{0.8\textwidth}
\setlength{\baselineskip}{0.9\baselineskip}
\footnotesize \textit{Notes:} Random effects logit regression coefficients are reported. Dependent variable is the channel choice (Channel B $= 1$). Standard errors are clustered at subject level. Decision set number and decision number within the decision set are controlled for. The following demographic variables are controlled for: age, gender, number of quiz errors and the Eckel-Grossman risk aversion measure (administered after the main task). H2 tests are one-sided, with $p-$values adjusted for multiple hypothesis testing using Bonferroni-Holm procedure. The remaining tests are two-sided. *** $p<0.01$ ** $p<0.05$ * $p<0.1$.
\end{minipage}
\end{table}

\subsubsection{Discussion}
In Experiment 2, we tested the effectiveness of two levers that managers can use to increase chatbot uptake: providing transparency about the chatbot's capabilities and highlighting potential time savings offered by the chatbot channel. Both remedies yield positive results. However, the effectiveness of each intervention varies with the duration of the interactions. When waiting times are short, participants are already choosing the chatbot relatively frequently, reducing the incremental benefit of emphasizing time savings. Under these shorter time horizons, participants appear more attuned to how the chatbot operates. Thus, being transparent about what the chatbot can and cannot do helps increase adoption. In contrast, when waiting times are long, participants are more reluctant to select the chatbot. In this scenario, highlighting the time-saving advantage is more compelling than additional transparency about capabilities. More broadly, the observed behaviors suggest that when the consequences of decisions are small, decision-makers focus on channel presentation and appearance of the channels. However, as waiting times increase and stakes are higher, decision-makers shift attention away from interactive features and toward efficiency considerations. Therefore, the relative effectiveness of different channel designs may depend on the time horizon. %We next leverage these experimental results to perform counterfactual analyses of channel selection behavior and quantify the resulting staffing cost savings for a service provider implementing chatbot technology.

\subsection{Structural Estimation of Staffing Cost Savings}\label{sec:structural}
To quantify the potential operational benefits of our experimental manipulations (queue time durations, transparency, nudge) we focus on staffing -- a key driver of controllable costs that motivated the deployment of customer service chatbots in the first place. We do this in three steps. We first formulate and estimate a random utility model of the customer choice between the live agent and chatbot channels. We then use the model estimates to predict customer arrival rates to each channel under various conditions. Finally, we calculate staffing costs based on the staffing levels necessary to maintain promised waiting times in each channel. This approach allows us to evaluate a range of counterfactual scenarios and estimate potential cost savings by explicitly accounting for endogenous customer channel selection.

\subsubsection{Structural Estimation of Utility Parameters}
To accurately predict channel arrival rates, we first need to estimate a customer utility function. In EC.4, we consider several plausible candidate utility functions. As is typical in structural estimation, increasing the number of parameters generally improves model fit but can reduce interpretability and intuition. Balancing this trade-off, we ultimately select a model that captures the aversion to using Channel B through a simple linear specification: 
\begin{align}
U^{A}_{ij}(\bm{\theta}) & = r - c_{line}\cdot t^A_{{line}_{ij}} - c_{agent}\cdot t^A_{{serve_1}_{ij}} + \epsilon^{A}_{ij},\\
U^{B}_{ij}(\bm{\theta}) & = r - c_{bot} \cdot t^B_{{serve_1}_{ij}} - (1 - p^B_{ij}) \cdot (c_{nt}  + \beta \cdot (c_{line}\cdot t^B_{{line}_{ij}} + c_{agent} \cdot t^B_{{serve_2}_{ij}})) + \epsilon^B_{ij},
\end{align} 
\noindent where $U^{A}_{ij}(\bm{\theta})$ and $U^{B}_{ij}(\bm{\theta})$ represent the utilities from receiving service through Channel A and Channel B, respectively, for participant $i$ in decision $j$. The parameter vector $\bm{\theta}$ includes the reward for service ($r$), waiting costs per second in line, with the agent, and with the chatbot ($c_{line}, c_{agent}, c_{bot}$), a lump-sum disutility if the chatbot fails and lacks transparency about its capabilities ($c_{nt}$), and a multiplier applied to the disutility from delays when the chatbot fails ($\beta$). This structure allows us to explicitly model how customers weigh different service channels based on expected waiting times and information availability. We normalize $r$ to zero and estimate $\bm{\theta}$ using maximum likelihood across the three treatments (\textit{Context, Context + Nudge, Context + No Transparency}), setting $c_{nt}$ to 0 in the \textit{Context} and \textit{Context + Nudge} treatments and estimating distinct $\beta$ parameters for treatments with and without the nudge. All parameters are statistically significant, with estimates and bootstrapped standard errors reported in EC.4.

\subsubsection{Demand Estimation}
To calculate staffing levels and costs for our counterfactual scenarios we need to know the arrival rates to each channel (demand intensities $\lambda^A$ and $\lambda^B$). We make the standard assumption that customers arrive according to a Poisson arrival process, with demand intensity $\lambda$, which then splits into $\lambda^A$ and $\lambda^B$ based on the offered waiting times and success probabilities. In particular, the relative demand for each channel is formed according to the logit choice probabilities $\rho^A(\bm{\theta})$ and $\rho^B(\bm{\theta})$, which can be derived from equations (4.1) - (4.2) and the estimates in Table EC.9. Consistent with the experimental setting, we model the live server as having deterministic service time. 

Despite these simplifications, characterizing the system entails solving for an equilibrium in which the waiting times initially promised ($t^A_{line}, t^B_{line}$) match the actual average waiting times ($\bar{t}_{line}$) arising from the endogenously determined arrival rates resulting from customer channel choices ($\rho^A(\bm{\theta}), \rho^B(\bm{\theta})$). We provide an intuitive overview of this procedure. We first compute the choice probabilities ($\rho^A(\bm{\theta}), \rho^B(\bm{\theta})$) for a range of counterfactual parameters, i.e., $t^A_{line},  t^B_{line}$, the presence (or absence) of the nudge and/or transparency, as well as the remaining system parameters. We then use these probabilities to calculate channel demand, where bot demand ($\lambda^B$) is simply the portion of total demand intensity $\lambda$ that is directed to the chatbot, and live agent demand ($\lambda^A$) is made up of two components: the customers who choose the live agent channel and the customers who choose the chatbot channel but experience chatbot failure and are redirected to the live agent. Finally, by modeling the system as an $M/D/1$ queuing regime, we are able to derive the live agent service rate $\mu$ required to deliver the announced waiting times.\footnote{We first compute the average sojourn time in the live agent channel, $T^A(\bm{\theta})=\bar{t}_{line}(\bm{\theta}) + t^A_{serve}$, where $\bar{t}_{line}(\bm{\theta})$ is the average time that a customer spends in line waiting for the live agent, weighted by the proportion of the live agent demand coming from each channel. Then the live agent service rate $\mu$ required to deliver the announced waiting times can be calculated as follows \citep[derived from][p. 59]{tijms2003}: $\mu(\lambda^A(\bm{\theta}),T^A(\bm{\theta})) = \frac{\lambda^A(\bm{\theta}) + \sqrt{\lambda^A(\bm{\theta})^2 + \frac{2\cdot \lambda^A(\bm{\theta})}{T^A(\bm{\theta})}}}{2}.$ The results are virtually identical if we use a $M/M/s$ system to model staffing costs.} If we assume that staffing costs increase linearly in the service rate $\mu$ (proxy for staffing level), we can use  our derivation to estimate staffing costs.

\subsubsection{Results}
We estimate staffing costs in five scenarios. In the baseline service design the chatbot is not transparent, no nudge is applied and the queue for the live agent is pooled (where $t^A_{line} = t^B_{line}$). This is compared against four scenarios that apply either transparency, the nudge, a priority queue for chatbot users (where $t^A_{line}$ is set lower than $t^B_{line}$)\footnote{To compare staffing costs with and without a priority queue for chatbot customers, we set $\bar{t}_{line}$ to be the same under each comparison between baseline and prioritization, but set $t^A_{line}$ and $t^B_{line}$ under prioritization such that the weighted average was equal to $\bar{t}_{line}$ and that $t^B_{line} = 0.9 \cdot t^A_{line}$. In other words, if the chatbot fails, the customer receives a 10 percentage point bump in the queue relative to a customer who immediately chose the agent channel. Giving strict priority to chatbot users would result in factors even lower than our chosen factor of 0.9. We chose a fairly conservative priority factor and held it constant across all scenarios to make like comparisons.}, or all three interventions combined. Figure 6 shows the relative staffing cost savings for counterfactual scenarios  defined by $\bar{t}^{line} \in [1, 200]$, and by $p^B = \{0.4, 0.5, 0.6\}$.\footnote{Further assumptions on system parameters are as follows. We set $\lambda$ to 0.1, resulting in a system utilization between 75\% and 80\% -- a utilization level commonly used in queuing analysis of moderate-to-heavy traffic. We set all service times to 20. Finally, we set the unit staffing cost to 1 (i.e., the staffing cost is simply given by the equation in the previous footnote).}

Figure 6 offers several managerial insights. First, in all scenarios the cost savings peak at an interior value of $\bar{t}_{line}$. When $\bar{t}_{line}$ is sufficiently low, as would be the case in a low-congestion system, joining the live agent queue for quick, guaranteed resolution is so attractive that any intervention has little effect on demand, explaining the low cost savings. Conversely, when $\bar{t}_{line}$ is high, as would be the case in a highly congested system, any intervention has little effect as the bot is already perceived as an attractive option to avoid the long wait for the live server. It is only for intermediate values of $\bar{t}_{line}$ that the interventions have a strong enough effect on demand to substantially decrease staffing costs. Second, among the three interventions, prioritizing chatbot customers for access to live agents has the highest potential to reduce staffing needs, yielding cost savings of between 11.5\% (in panel c) and 14.7\% (in panel a). This suggests that operational efficiency gains from priority queues exceed those from information-based interventions. Third, the highest cost savings occur for chatbots of intermediate capability ($p^B=0.5$), with up to $19.7\%$ cost savings, suggesting that design interventions have their greatest impact when chatbots are neither too ineffective nor too advanced, representing an important transition zone where customer channel choice is most malleable.

\begin{figure}[tbh!]
\centering
\caption{Counterfactuals: Staffing Cost Savings}
\label{fig:cost:savings}
\includegraphics[width=\textwidth]{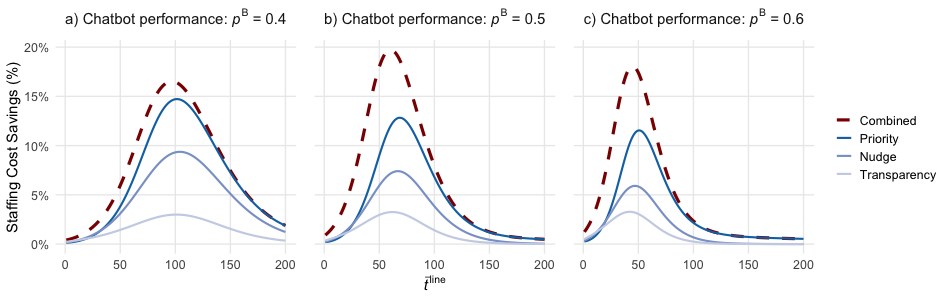}
\end{figure}

\section{Experiment 3: Alternative Measurements of Algorithm Aversion}\label{sec:exp:3}
In Experiments 1 and 2 we examined key drivers of chatbot adoption and identified ways to overcome barriers to use. In both of those experiments, participants interacted with servers through identical click-based prompts across channels. This enabled precise control and measurement of interaction times, ensuring that the actual time spent matched the time promised when participants made their channel selections. Although this design provided strong experimental control, it necessarily reduced interaction realism. Recognizing the trade-off between experimental control and realism, in Experiment 3 we replicate  our original design, but add more authentic representations of how users may experience interactions across different service channels.

\subsection{Methodology}
\subsubsection{Treatments}
Experiment 3 consisted of two treatments, which we will refer to as the \textit{Context + Live} treatment and the \textit{Context + Hold} treatment. A total of 222 new subjects were recruited to participate in these treatments of whom 177 passed all comprehension and consistency checks. The training materials for the research assistants representing the live agents are in EC.3.3. As in Experiments 1 and 2, participants made 33 decisions, with three of these decisions being implemented after the decisions were submitted. As in Experiments 1 and 2, participants were shown a demo of each channel prior to making their decisions. Both treatments were conducted using the short duration parametrization.
 
The \textit{Context + Live} treatment was identical to the \textit{Context} treatment in Experiment 1, with the only difference being that the role of the live agent was now played by an experimenter. To this end, we recruited two research assistants who had no prior knowledge of the hypotheses. We followed common practices for deploying confederates (research assistants) in experimental research \citep{kuhlen2013} and trained the assistants using a script to control for potential differences in communication patterns (See EC.3.3 for the script). At the beginning of the service process, the participant indicated their assigned ``issue type'', after which the research assistant started the service process, which then lasted for the required duration ($t^B_{serve_2}$ seconds).%\footnote{To ensure that participants' beliefs about the (human/computerized)  nature of the server did not influence their behavior, at the end of the experiment we asked participants whether they thought that they were interacting with a human or with a computer in each channel. The results show that 63\% of participants believed they were interacting with a human agent in Channel A (compared to only 12\% who believed this about the Channel B), suggesting that our manipulations worked as intended for the majority of participants.} 

The \textit{Context + Hold} treatment was identical to the \textit{Context} treatment in Experiment 1, with the only difference being that the interaction mode differed between the channels. In particular, we made the human/algorithmic nature of the server more salient. In interacting with the server representing the live agent, participants were required to hold down a button for the duration of service. In contrast, in interacting with the server representing the chatbot, participants continued using keystrokes to interact, exactly like in Experiment 1. See EC.3.3 for the description of the experimental stimuli.\footnote{The hold vs. click-based interaction modes were chosen to represent a more continuous interaction with an agent vs. a more fragmented interaction with a bot. To ensure that these interaction modalities were valid representations of the channels, we performed a manipulation check (with a separate group of respondents), which confirmed that the manipulations were associated with the channel as intended. See Appendix EC.3.3 for details.}

\subsubsection{Theory and Hypotheses}
In Experiments 1 and 2, we followed an approach common in the experimental literature and examined how contextual information, i.e., variations in instructions and visuals, affects chatbot uptake. However, context is only one part of understanding algorithm aversion, particularly in customer service, where \textit{experiential} factors also play a role. The interventions in Experiment 3 are designed to amplify these experiential differences between channels.  In particular, the visible presence of a human may increase algorithm aversion by creating a stronger contrast and making the alternative (Chatbot channel) appear more algorithmic. Similarly, introducing physical engagement, such as holding a button rather than clicking prompts in Channel A, may increase the perception of interacting with a human by simulating agency and control over the interaction. Thus, we hypothesize:

\noindent\textbf{H3.1 (Live):} Conditional on expected waiting times, participants in the \textit{Context + Live} treatment choose the Chatbot channel less frequently than in the \textit{Context} treatment. \newline
\noindent\textbf{H3.2 (Hold):} Conditional on expected waiting times, participants in the \textit{Context + Hold} treatment choose the Chatbot channel less frequently than in the \textit{Context} treatment.

\subsection{Results}
\subsubsection{Descriptive Statistics}
Figure 7 shows average chatbot uptake by decision, with the \textit{Context} treatment added as a comparison baseline. The figure suggests that chatbot uptake goes down in the \textit{Context + Live} treatment relative to the \textit{Context} treatment, with decreases between two and eight percentage points, depending on the decision. Further, chatbot uptake also goes down in the \textit{Context + Hold} treatment, with decreases between two and seventeen percentage points. Thus, both manipulations aimed at making the perceptual differences between channels more salient appear to increase algorithm aversion, providing some initial support for H3.1 and H3.2. 

\begin{figure}[b!]
\centering
\caption{Channel B (Chatbot Channel) Uptake in Experiment 3}
\label{fig:bars_s3}
\includegraphics[width=0.7\textwidth]{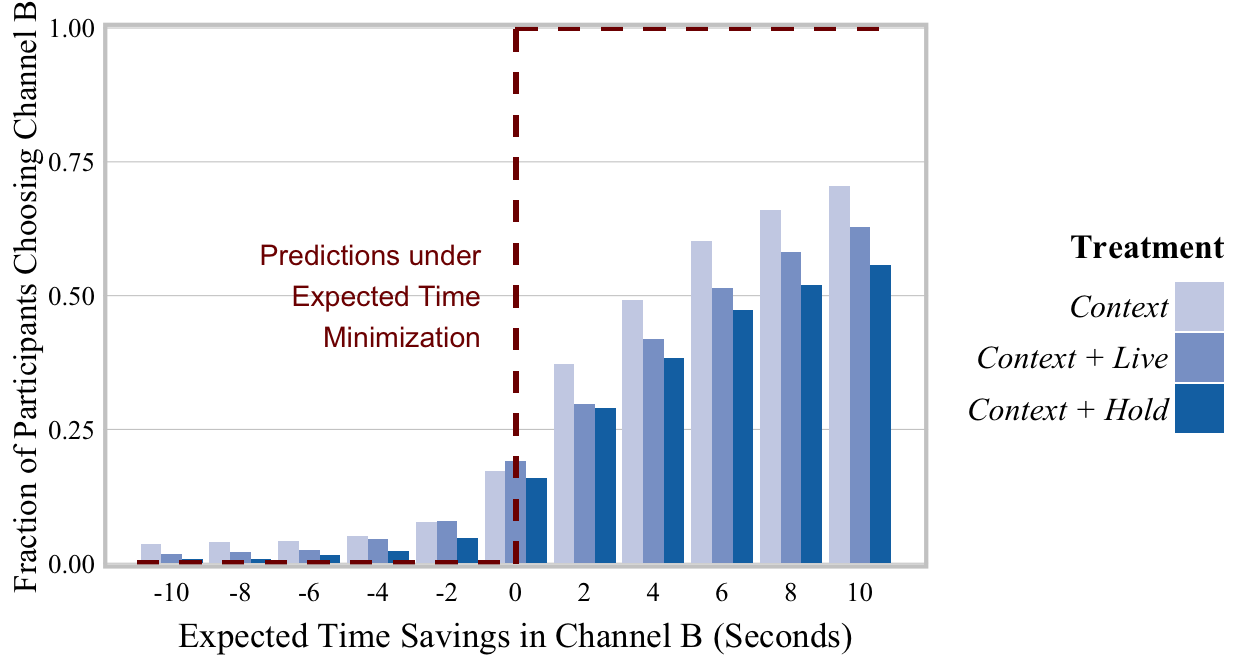}
\end{figure}
 
\subsubsection{Hypothesis Tests}
To test H3 we regressed channel choice on the treatment dummies. Table \ref{tab:reg:exp3} presents the estimates. As before, in Table \ref{tab:reg:exp3}  we report one-sided, Bonferroni-Holm adjusted $p-$values. Table \ref{tab:reg:exp3} shows that both the effects of introducing perceptual differences between channels (through manipulating how the user interacts with the server), and introducing a live human into the interaction, reduces chatbot uptake. In particular, adding a live human to the service process decreased chatbot uptake by 7.02 percentage points, on average ($p=0.013$). Similarly, introducing server-specific interaction modes (holding down a button for live agent vs. prompt clicking for chatbot) decreased chatbot uptake by 9.92 percentage points, on average ($p<0.01$). Thus, the regression results confirm that the observed increase in algorithm aversion is indeed statistically significant.\footnote{In post-hoc comparisons we find that the difference between \textit{Context + Live} and \textit{Context + Hold} treatment dummies is not statistically insignificant ($p=0.225$), suggesting that the two manipulations led to similar levels of algorithmic aversion. }

\begin{table}[tb!]
\renewcommand{\arraystretch}{1.1}
\caption{Channel Preferences in Experiment 3}
\label{tab:reg:exp3}
\footnotesize
\centering
\begin{tabular}{lC{8cm}}
\toprule
& (1) \\
\multicolumn{1}{r}{Dependent Variable:} & Channel B (Chatbot Channel) \\
\midrule
\textit{Context} Treatment & Omitted category \\
\addlinespace
\textit{Context + Live} Treatment & -1.064** \\
& (0.536) \\
\addlinespace
\textit{Context + Hold} Treatment & -1.737*** \\
& (0.540) \\
%\addlinespace
%Constant & -4.593*** \\
%& (1.111) \\
\midrule
Channel Performance Controls & Yes \\
Demographic Controls & Yes \\
Sample & Full Sample \\
Observations & 9240 \\
Subjects & 280 \\
\bottomrule
\end{tabular}

\vspace{0.3em}
\begin{minipage}{0.8\textwidth}
\setlength{\baselineskip}{0.9\baselineskip}
\footnotesize \textit{Notes:} Random effects logit regression coefficients are reported. Dependent variable is the channel choice (Channel B $= 1$). Standard errors are clustered at subject level. Decision set number and decision number within the decision set are controlled for. The following demographic variables are controlled for: age, gender, number of quiz errors and the Eckel-Grossman risk aversion measure (administered after the main task). H3 tests are one-sided, with $p-$values adjusted for multiple hypothesis testing using Bonferroni-Holm procedure. The remaining tests are two-sided. *** $p<0.01$ ** $p<0.05$ * $p<0.1$.
\end{minipage}
\end{table}
\smallskip
\noindent \textit{\textbf{Result 3 (Algorithm Aversion):} H3.1 and H3.2 are supported. Channel B uptake is reduced under the \textit{Context + Live} and the \textit{Context + Hold} manipulations relative to the Context treatment.}

\subsection{Discussion}
In Experiment 3 we explored the trade-off between experimental control and external validity when studying chatbot adoption in service settings. Adding realistic elements such as live interactions and physical engagement creates a more authentic experience but can potentially reduce experimental control, making it more difficult to measure exact times spent or rule out alternative explanations. Despite these more realistic implementations, the overall magnitude of algorithm aversion in Experiment 3 was comparable to that in Experiment 1, suggesting that our key results are robust to alternative implementations. At the same time, we saw that perceptual differences contribute modest (though statistically significant) effects to the overall aversion. This has both practical and methodological implications. From the practical standpoint, it suggests that the effects of the remedies in Experiment 2 (priority queues, transparency and reducing perceived uncertainty through nudges) likely present a lower bound on potential cost savings. Indeed, the stronger aversion levels observed in Experiment 3 provide more room for improvement, which may lead to even greater savings from improvements in service design. Lastly, from a methodological perspective, the results presented in this section suggest that future research on algorithmic technology adoption should exercise caution when using purely contextual, framing-based manipulations, as these can significantly underestimate algorithm aversion.

\section{Concluding Remarks}\label{sec:conclusions}
AI-powered chatbots are becoming an increasingly integral part of online customer service. To successfully leverage chatbot technology, firms need to understand both the relevant customer choice trade-offs and their operational implications. In this paper we studied chatbot adoption by first soliciting and analyzing testimonies from chatbot users, by using these user stories to formulate a key trade-off in channel choice, by examining how users navigate this trade-off in incentivized experiments, and by evaluating the cost savings achievable through simple process-design interventions.

\subsubsection*{Summary of Results}
The standard approach in the service operations literature is to model queue-joining behavior as a simple expected-time minimization problem \citep{naor1969,hassin2003}. Our experimental results suggest that this approach may oversimplify behavior when one service channel has a gatekeeper structure. Specifically, we found that most decision-makers avoided the gatekeeper channel even when this channel offered shorter expected waiting times. We termed this behavior \textit{gatekeeper aversion} and characterized is as a combination of risk and transfer aversion. Separately, we identified another hurdle to chatbot uptake -- \textit{algorithm aversion} -- and showed that it can amplify gatekeeper aversion. In particular, while algorithm aversion was not detectable in isolation, i.e., in our pre-tests where the service processes and performance were identical across channels, associating the gatekeeper channel with the chatbot further decreased participants’ willingness to use it. Lastly, we examined potential remedies for chatbot underutilization and showed that priority queues, operational transparency and an average waiting time nudge increase adoption. Using counterfactual analysis of channel joining behavior in an $M/D/1$ system, we showed that, when combined, these remedies can yield staffing cost savings of up to 19.7\%.

\subsubsection*{Contributions}
We make several contributions to theory, practice, and experimental methodology.  On the theory front, we extend the conversation on human-AI interfaces, which has traditionally focused on AI adoption among workers \citep{dietvorst2015,yeomans2019,jussupow2020}, to customer decisions in service channel selection. By connecting this research stream to the service operations literature on queue-joining behavior \citep{allon2018} and gatekeeper service systems \citep{shumsky2003,freeman2017}, we show that the classic finding that errors loom larger when made by algorithms does not seamlessly generalize to our customer service context. Instead, algorithm aversion plays a more subtle role: it amplifies existing reluctance to engage with a channel that has a gatekeeper structure. The implication of this result is that the reluctance to using chatbot technology is likely to drop substantially once this technology reaches performance levels similar to humans. Second, we make a more practical contribution by showing that low-cost, easily implementable service design interventions -- such as introducing priority queues, providing average wait information, and truthfully revealing chatbot capabilities -- can significantly increase chatbot adoption and generate substantial cost savings for the service provider. Third, we develop a novel experimental approach for eliciting algorithm aversion in service systems and show that experimental designs relying solely on contextual framing may underestimate algorithm aversion, compared to designs that vary the human vs. algorithmic nature of the interaction in a more realistic manner.

%\subsubsection*{Service Design Implications}
%Our findings offer new practical insights for managers. The positive response to chatbot performance parameters suggests that managers should continue investing in chatbot performance. Such investments can target improvements in chatbot reliability and range, or reductions in the time needed to correctly diagnose and respond to customer requests. At the same time, our data suggest that the returns to improvements in performance may be marginally decreasing in certain ranges -- a nontrivial share of decision-makers in our data avoided the chatbot channel regardless of its performance. Second, our results suggest a simple operational lever that can increase chatbot adoption: prioritizing chatbot customers for quicker access to a live agent when the chatbot fails to resolve their issue. Because waiting in line after chatbot failure is especially painful, reducing these waits will prompt more customers to opt for the chatbot channel. The shift  towards the chatbot channel can reduce the demand for live agents and free up staffing capacity, which can then be used to cut staffing or to improve service quality in the live agent channel. Third, our results on algorithm aversion suggest that managers need to be cognizant of age as a significant moderator of chatbot uptake and may need to customize the available channels to the type of product and target demographics.  
 
\subsubsection*{Limitations and Extension}
To keep the experiments focused, we did not model certain aspects of channel choice, such as the language and style of service interactions, the seamlessness of transitions between gatekeepers and experts, the uncertainty in interaction times, or the residual probability of expert failure. Examining these features may add realism to our setup and may improve the generalizability of our findings. Avenues for future research also include the role of algorithmic preferences when interacting with firm-specific vs. general-purpose chatbots (e.g., ChatGPT), the use of chatbots to perform purely diagnostic work vs. more task-oriented functions, and the role of privacy concerns and the use of customer data in these service interactions.  Finally, our experiments in \textsection 5, which tested basic variations in interaction mode, suggest that richer server-customer interaction environments such as voice, video, or virtual reality deserve further study as they could produce markedly different levels of algorithm aversion.

\subsubsection*{Outlook}
The customer service context provides an ideal setting for our study because it allows precise control and communication of waiting times, involves outcomes that are binary (a request is either resolved or not), and is largely conducted online rather than in physical retail stores. Therefore, our online experiments mirror service interactions with high fidelity to the real-world setting. However, given that similar technology adoption challenges exist beyond customer service, we believe that some of our results may apply more broadly. Examples include self-checkout stations in grocery stores, automated check-in kiosks at airports, or digital ordering systems in restaurants. Future research could examine whether key behaviors identified in chatbot interactions -- such as gatekeeper aversion and algorithm aversion -- are also observed in these alternative automated settings, and how service design interventions like operational transparency and pooled/dedicated queues might mitigate technology underutilization more broadly. As self-service technologies continue to evolve and mature, controlled experiments offer a powerful tool that can add to our understanding of customer behavior and service design. 

%\newpage \clearpage \raggedbottom 
\bibliographystyle{informs2014} % outcomment this and next line in Case 1
\bibliography{literature} % if more than one, comma separated

\end{document}